\documentstyle[pre,aps,multicol,epsf]{revtex}
\def\gs{> \kern -12pt \lower 5pt \hbox{$\displaystyle{\sim}$}}
\def\ls{< \kern -12pt \lower 5pt \hbox{$\displaystyle{\sim}$}}
\def\gdot{\dot{\gamma}}
\def\be{\begin{equation}}
\def\en{\end{equation}}
\newcommand{\bi}[1]{\mbox{\boldmath$#1$}}
\renewcommand{\theequation}{\arabic{section}.\arabic{equation}}
\newcommand{\av}[1]{\langle{#1}\rangle}
\def\bea{\begin{eqnarray}}
\def\ena{\end{eqnarray}}
\def\p{\partial}
\def\Ge{\Gamma_{eff}}

\begin{document}
\draft
\title{Dynamics of Highly Supercooled Liquids:\\
Heterogeneity, Rheology, and Diffusion}

\author{Ryoichi Yamamoto and Akira Onuki}
\address{
Department of Physics, Kyoto University, Kyoto 606-01
\\ E-mail: ryoichi@ton.scphys.kyoto-u.ac.jp, onuki@ton.scphys.kyoto-u.ac.jp
}

\date{\today}
\maketitle
\begin{abstract}

Highly supercooled liquids
with soft-core potentials are studied
via molecular dynamics simulations in two and three dimensions
in quiescent and sheared conditions.
We may define bonds between  neighboring
particle pairs unambiguously owing to the sharpness
 of the first peak of the pair correlation functions.
Upon structural rearrangements,  they
break  collectively in the form of clusters
whose  sizes grow  with lowering the temperature $T$.
The bond life time $\tau_b$, which depends  on
$T$ and the shear rate
$\gdot$, is on  the order of the usual
structural or $\alpha$ relaxation time $\tau_{\alpha}$
in weak shear $\gdot \tau_{\alpha} \ll 1$,
while  it decreases as $1/\gdot$ in
strong shear $\gdot\tau_{\alpha} \gg 1$ due to
shear-induced cage breakage.
Accumulated broken bonds
in a  time interval ($\sim 0.05\tau_b$)
 closely resemble the
 critical fluctuations of Ising spin systems.
For example, their  structure factor  is
well fitted to the
Ornstein-Zernike form,
which yields the correlation length $\xi$ representing
the maximum size of the clusters composed of broken bonds.
We also find a dynamical scaling
relation,  $\tau_b \sim \xi^{z}$,  valid for any $T$
and $\gdot$   with $z=4$ in two
dimensions and $z=2$ in three dimensions.
The viscosity is of order $\tau_b$ for any $T$
and $\gdot$,  so  marked  shear-thinning
behavior emerges. The shear stress
is close  to a limiting stress in a wide shear region.
We also  examine motion  of tagged particles in shear
in three dimensions.  The diffusion constant is
found to be of order $\tau_b^{-\nu}$
with $\nu=0.75 \sim 0.8$ for any $T$
and $\gdot$,
so it is much enhanced in strong shear compared with
its value at zero shear.
This indicates  breakdown of the Einstein-Stokes relation
  in accord with experiments.  Some possible experiments
  are also proposed.
\end{abstract}

%\narrowtext
\begin{multicols}{2}
\section{Introduction}

Particle motions in supercooled liquids are severely restricted
or jammed, thus giving rise to slow structural relaxations and
highly viscoelastic behavior  \cite{Jackel,Ediger}.
Recently  much  attention has been paid to
the mode coupling theory, \cite{mode1,mode2}
 which  is a first analytic
  scheme  describing   onset of
 slow structural relaxations
 considerably above $T_g$.  There,
 the density fluctuations with
 wave numbers around
 the first peak position  of the structure factor
 are of most importance and
 no  long range correlations are predicted.
For a long time, however,
 it has been  expected
 \cite{Adam,Cohen,Stillinger,Ngai} that
rearrangements of particle configurations in glassy materials
 should be cooperative,
involving many molecules,  owing to configuration
 restrictions. In other words, such events
 occur only in the form of {\it clusters}
whose sizes increase at low temperatures.
In normal liquid states, on the contrary, they are frequent and
 uncorrelated among one another in space and time.
Such an idea was first put forth by Adam and Gibbs \cite{Adam},
who invented a frequently
used jargon, {\it cooperatively rearranging regions} (CRR).
However, it is difficult to judge whether or not such phenomenological
models are successful in describing real physics and in  making
quantitative predictions.

Molecular dynamics (MD) simulations can be powerful tools to
gain insights into relevant physical processes in highly
supercooled liquids.
Such processes are  often masked in
averaged quantities such as the
density time correlation functions.
As a marked example,
we mention kinetic heterogeneities observed
in recent simulations
\cite{Muranaka1,Muranaka2,Hurley,Perera,Yamamoto_Onuki1,Yamamoto_Onuki2,Vigo,Kob,Donati,Ray}.
Using  a simple two dimensional fluid,
 Muranaka and Hiwatari
\cite{Muranaka1} visualized significant
large scale heterogeneities in particle displacements in a
relatively short time interval which was supposed
to correspond to the $\beta$ relaxation time regime.
In liquid states with higher temperatures,
Hurley and Harrowell \cite{Hurley}
observed similar kinetic heterogeneities
but the correlation length was still
on the order of a few particle diameters.
The characterization of these patterns has not been made in
these papers.
Recently our simulations on model fluid mixtures
in two and three dimensions
\cite{Yamamoto_Onuki1,Yamamoto_Onuki2,Vigo}
have identified {\it weakly bonded}
or {\it relatively active} regions from
breakage of appropriately
defined bonds.
Spatial distributions of such regions resemble the critical
fluctuations in Ising spin systems,
so the  correlation length $\xi$ can be determined.
It grows up to the system size as $T$ is lowered, but no
divergence seems to exist
at nonzero temperatures
\cite{Yamamoto_Onuki1,Dasgupta,Ernst,Ghosh}.
Donati {\it et al.} have observed {\it string-like}  clusters
 whose lengths increase at low temperatures in a three dimensional
  binary mixture  \cite{Donati}. In addition,
Monte Carlo
simulations of a dense polymer
by Ray and Binder showed
a significant system size dependence
of the monomer diffusion constant, which indicates
heterogeneities over the system size  \cite{Ray}.

Most previous papers so far are concerned with near-equilibrium
properties such as  relaxation of the density time correlation
functions or dielectric response.
>From our point of view, these quantities
are too restricted or indirect, and
there remains a rich group of unexplored problems
in far-from-equilibrium  states.
For example, nonlinear glassy response
against electric field, strain, {\it etc.} constitutes
a future problem \cite{Osheroff}.
In this paper we apply a simple shear
flow ${v}_x= \gdot y$ in the $x$ direction and
realize  steady states \cite{Onuki_review}.
The velocity gradient
$\gdot$  in the $y$ direction is  called the shear rate
or simply shear.  We shall see that it
   is a relevant perturbation
drastically changing the glassy  dynamics when
$\gdot$ exceeds the inverse of the
structural or $\alpha$ relaxation time $\tau_{\alpha}$.
As is well known, $\tau_{\alpha}$ increases dramatically
from  microscopic to macroscopic times in a rather
narrow temperature range \cite{Jackel,Ediger}.
Generally, in near-critical fluids and
 various complex fluids,
 nonlinear shear regimes are known to
 emerge when $\gdot$ exceeds
 some underlying relaxation rate \cite{Onuki_review}.
However, in supercooled liquids,
it is  unique that
 even very small shear can greatly accelerate the
{\it microscopic}  rearrangement processes. Similar effects
are usually  expected  in   systems
composed of very large elements such as colloidal suspensions.

Though rheological
experiments on glass-forming fluids
 have not been abundant,
Simmons {\it et al} found that the viscosity
$\eta (\gdot) =\sigma_{xy}/\gdot$
exhibits strong shear-thinning behavior,
\be
\eta (\gdot) \cong \eta (0)/(1+\gdot \tau_{\eta} ),
\en
in  soda-lime-silica glasses in steady states under
shear \cite{Simons1,Simons2,Yue}, $\tau_{\eta}$
being a long rheological time.
After application of shear,  they also observed
overshoots of the shear stress before approach to steady states. 
Our previous reports \cite{Yamamoto_Onuki2,Vigo} have treated
nonlinear rheology in supercooled liquids in agreement with
these these  experiments.
Interestingly, similar {\it jamming dynamics}
has begun to be recognized also in rheology of
foams \cite{Okuzono,Durian,Liu} and granular materials \cite{Behringer}
composed of  large  elements.
Shear-thinning behavior and heterogeneities
in configuration rearrangements are commonly
observed also in these macroscopic systems.

As a closely  related problem,
understanding of
mechanical properties of amorphous metals such as
Cu$_{57}$Zr$_{43}$ has  been
of great technological importance
\cite{HSChen,Spaepen,Aragon,Takeuchi,Takeuchi_review}.
They are usually ductile in spite of their high strength.
At low temperatures $T \ls~ 0.6 \sim 0.7 T_g$
localized bands ($\ls~ 1 \mu m$),
where  zonal slip occurs,  have been observed
above a yield stress. At relatively high temperatures
$T \gs~ 0.6 \sim 0.7 T_g$, on the other hand,
shear deformations are induced
{\it homogeneously} (on macroscopic
scales) throughout
samples, giving rise to
 viscous flow with strong shear
thinning behavior. In particular, in their 3D simulations 
Takeuchi {\it et al.} \cite{Takeuchi,Takeuchi_review} followed atomic
motions after application of a small shear strain to observe
 heterogeneities among {\it poorly and closely
 packed regions},
which are essentially the same entities we have
discussed. Our simulations under shear in this paper
correspond to the {\it homogeneous} regime at relatively high
temperatures  in amorphous metals.

Another interesting issue is as follows.
Several experiments have revealed  that
the translational diffusion constant $D$ of a tagged particle
in a fragile glassy matrix becomes increasingly larger
than the Einstein-Stokes value
 $D_{ES}=  k_BT/2\pi \eta a$ with lowering $T$,
where $\eta$ is the (zero-shear) viscosity and $a$ is
the diameter of the particle \cite{Ediger,Silescu,Ci95,St94}.
In particular, the power law behavior
$D \propto \eta^{-\nu}$ with $\nu \cong 0.75$
was observed at sufficiently low temperatures \cite{Silescu}.
Thus $D/D_{ES}$ increases from of order 1
up to order $10^2 \sim 10^3$ in supercooling experiments.
Furthermore, smaller  probe particles
exhibit a more pronounced increase of
the ratio $D\eta/T \propto D/D_{ES}$ with lowering $T$  \cite{Ci95}.
It is generally believed that $\eta$ is proportional to
the $\alpha$ relaxation time $\tau_{\alpha}$ or the rotational
relaxation time $1/D_{rot}$ for anisotropic molecules
($D_{rot}$ being the rotational diffusion constant) 
\cite{Silescu,Ci95,Richter}. Therefore,  individual particles
 are much more mobile at long times
$t \gs~ \tau_{\alpha}$ than
expected from the Stokes-Einstein
formula.
In a  MD simulation on  a 3D  binary mixture
with $N=500$ in 3D  \cite{Mountain}
the same tendency was
apparently seen despite of their small system size.
Very recently, in a MD simulation in a 2D binary
mixture with $N=1024$,  Perera and Harrowell have observed
clear deviation from  the linear  relation
$D \propto \tau_{\alpha}$ where $\tau_{\alpha}$
is obtained from the decay of the time correlation function
as in our case in Section 6
\cite{Perera_private}.
We will examine this problem  in a much larger 3D system
with $N=N_1+N_2=10^4$ generally in the presence of shear,
where  the viscosity and the diffusion constant
 both vary   tremendously in
 strong shear  ($\gdot \gs~ 1/\tau_{\alpha}$).

The organization of this paper is as follows.
In Sec.II  our  model binary mixtures
 and  our simulation method will be explained.
 In Sec.III
 {\it bonds} among particle
 pairs will be introduced at distances
 close to  the first peak position  of
  the pair correlation functions.
Breakage of such bonds will then be followed numerically,
which exhibits   heterogeneities enhanced at low temperatures.   
Their  analysis will yield the correlation length
in Sec. IV. Rheology of supercooled liquids will
be studied in Sec.V.
These effects were
 briefly reported in our previous reports
\cite{Yamamoto_Onuki1,Yamamoto_Onuki2,Vigo}.
In Sec.VI new results on motion  of tagged particles
will be presented.

\section{Model and simulation method}
\setcounter{equation}{0}

We performed  MD simulations
 in two dimensions (2D) and three
dimensions (3D) on binary mixtures composed of two different
atomic species, $1$ and $2$, with $N_{1}=N_{2}=5000$ particles
 with the system volume $V$ being fixed.
Parameters chosen are mostly common in 2D and 3D.
They interact via the soft-core potential
\cite{Muranaka1,Muranaka2,Hurley,Perera,Yamamoto_Onuki1,Yamamoto_Onuki2,Vigo,Perera_private,Matsui,Bernu},
\be
v_{\alpha\beta}(r)=
\epsilon (\sigma_{\alpha\beta}/r)^{12} ,
\qquad
\sigma_{\alpha\beta}=\frac{1}{2}
(\sigma_{\alpha}+\sigma_{\beta}),
\label{eq:2.1}
\en
where $r$ is the distance between two particles and
$\alpha, \beta= 1,2$.
The interaction is truncated at $r =4.5\sigma_{1}$ in 2D and
$r =3\sigma_{1}$ in 3D.
The leapfrog algorithm is used  to integrate  the differential
equations with a time step of $0.005\tau_0$, where
\be
\tau_0=({m_{1}\sigma_{1}^{2}/\epsilon})^{1/2}.
\label{eq:2.2}
\en
The space and time are measured in units of $\sigma_1$ and
 $\tau_0$. The  mass ratio is  $m_{2}/m_{1}=2$, while  the size ratio is
\be
\sigma_{2}/\sigma_{1}=1.4 \quad ({\rm d}=2), \quad
\sigma_{2}/\sigma_{1}=1.2  \quad ({\rm d}=3),
\label{eq:2.3}
\en
where $d$ is the space dimensionality.
This  size  difference  prevents crystallization
 and produces amorphous states in our systems
at low temperatures.

We fixed  the particle density  at
\be
n=0.8/\sigma_{1}^{d},
\label{eq:2.4}
\en
where $n=n_1+n_2$ is the total number density.
The system linear
dimension is $L=118$ in 2D and $L=23.2$   in 3D.  Then
 our systems are highly compressed.
In fact, the volume fraction of the particles
may be estimated as
$\pi (\sigma_1^2 n_1+ \sigma_2^2n_2 )= 0.93$ in 2D and
as  $\frac{4}{3}\pi
(\sigma_1^3 n_1+ \sigma_2^3n_2 )= 0.57$ in 3D, where overlapped
regions are doubly counted.
In such cases, according to the Henderson and Leonard theory
\cite{Bernu,Henderson,Hansen},
 our binary mixtures   may be fairly
mapped onto one-component fluids with the soft-core potential
 with an  effective
radius defined by
\be
\sigma_{eff}^d = \sum_{\alpha,\beta=1,2} x_{\alpha}x_{\beta}
\sigma_{\alpha\beta}^d ,
\label{eq:2.5}
\en
where $x_1=n_1/n$ and $x_2=n_2/n=1-x_1$ are the
compositions of the two components
and are $1/2$ in our case.
As in the one component case the thermodynamic
 state is  characterized by
 a single parameter (effective density),
\be
\Gamma_{eff}= n(\epsilon/k_BT)^{d/12}
\sigma_{eff}^d .
\label{eq:2.6}
\en
For example, Bernu {\it et al.} \cite{Bernu}
confirmed  that   the equilibrium pressure $p$ may be
well fitted to  the scaling form, $p/nk_BT-1 \cong  6+ 6.848(\Gamma_{eff})^4$,
at all $x_1$ in 3D.
In Tables I and II  we list $\Gamma_{eff}$ chosen
 in our simulations  together with  the corresponding
 scaled temperatures  and pressures in 2D and 3D. Our  pressure data
 excellently agree with the above scaling form for  3D.
%%%%%%%%%  Tables of parameters should appear here %%%%%%%%%%%%%%%%%%

We here introduce  the pair correlation
functions $g_{\alpha\beta}(r)$ by
\be
\av{ {\hat{n}}_{\alpha}({\bi r})
{\hat{n}}_{\beta}({\bi 0})} = {n_{\alpha}n_{\beta}} g_{\alpha\beta}(r)
+ {n_{\alpha}}\delta_{\alpha\beta}\delta ({\bi r}),
\label{eq:2.7}
\en
where
\be
\hat{n}_{\alpha}({\bi r})=
\sum_{j}\delta ({\bi r}-{\bi r}_{\alpha j}) \qquad
 (\alpha=1,2)
\label{eq:2.8}
 \en
are the number densities in terms of the particle positions
${\bi r}_{\alpha j}$ $(\alpha=1,2, j=1,\cdots, N/2$).
The time dependence is suppressed for simplicity.
In a highly compressed state the inter-particle
distances between
the $\alpha$ and $\beta$ particles are characterized by
\be
\ell_{\alpha\beta}= \sigma_{\alpha\beta}(\epsilon/k_BT)^{1/12}.
\label{eq:2.9}
\en
The last factor $(\epsilon/k_BT)^{1/12}$
represents the degree of expulsion or  penetration from or  into
the soft-core regions ($r < \sigma_{\alpha\beta}$) on particle encounters,
though it is not far from 1 in our case.
The one-fluid approximation
may be justified if the pair correlation functions  satisfy
\be
g_{\alpha\beta}(r)= G(r/\ell_{\alpha\beta}, \Gamma_{eff}) .
\label{eq:2.10}
\en
Namely, $g_{\alpha\beta}(r)$ are  independent of $\alpha$, $\beta$,
and $x_1$ once the distance is  scaled by $\ell_{\alpha\beta}$.
The pressure is then expressed as \cite{Hansen}
\bea
\frac{p}{nk_BT}-1 &=& -\frac{n}{2dk_BT}\sum_{\alpha,\beta}
\int d{\bi r}x_{\alpha}x_{\beta}v_{\alpha\beta}'(r) 
r g_{\alpha\beta}(r) \nonumber \\
&=&  6V_d\Gamma_{eff}\int_0^{\infty} ds
\frac{1}{s^{13-d}}G(s, \Gamma_{eff}) ,
\label{eq:2.11}
\ena
where $v_{\alpha\beta}'(r)=dv_{\alpha\beta}(r)/dr$ and
$V_d$ is the volume of a unit sphere, so
it is $4\pi/3$ in  3D and $\pi$ in 2D.
We shall see that (2.10) excellently  holds around the first peak of the
pair correlation functions in our simulations.  This fairly supports
 the one-fluid approximation,  because   the
 soft-core potential and the pair correlation functions
decrease   very abruptly for $r \gs~ \ell_{\alpha\beta}$ and for
$r \ls~ \ell_{\alpha\beta}$,   respectively, and the dominant contribution
  arises from $r \sim \ell_{\alpha\beta} \sim \sigma_{\alpha\beta}$.

In our systems the structural relaxation time
becomes very long at  low temperatures.
Therefore, the  annealing time  was taken to be at least
$10^5$ in 2D and $10^4$  in 3D.
No appreciable aging effect was detected in
the course of taking data  in various quantities such as the pressure
or the density time correlation function
except for the lowest temperature cases,
 $\Gamma_{eff}=1.4$ in 2D and $\Gamma_{eff}=1.55$ in 3D.
A small aging effect remained in the density time correlation function
in  these  exceptional cases, however.

%%%%%%  Tables %%%%%%
\narrowtext
\begin{table}
\caption{Simulations in 2D.}
\label{tab:2d}
\begin{tabular}{ccccccc}
$\Gamma_{eff}$  && 1.0 & 1.1 & 1.2 & 1.3 & 1.4 \\
\hline
$k_BT/\epsilon$  && 2.54 & 1.43 & 0.85 & 0.526 & 0.337 \\
\hline
$p/nk_BT-1$  && 15.1 & 22.6 & 33.5 & 50.2 & 75.1 \\
\end{tabular}
\end{table}
\begin{table}
\caption{Simulations in 3D.}
\label{tab:3d}
\begin{tabular}{cccccccc}
$\Gamma_{eff}$  && 1.15 & 1.3 & 1.4 & 1.45 & 1.5 & 1.55 \\
\hline
$k_BT/\epsilon$  && 0.772 & 0.473 & 0.352 & 0.306 & 0.267 & 0.234 \\
\hline
$p/nk_BT-1$  && 18.9 & 26.7 & 33.4 & 37.2 & 41.4  & 46.3 \\
\end{tabular}
\end{table}

\vskip-0.5pc

Our simulations were performed in the absence and presence of 
shear flow \cite{Allen,Evans}.
In the  unsheared case   ($\gdot=0$)
we performed simulations under the microcanonical
(constant energy)
condition. However, in the sheared case ($\gdot > 0$), we
kept  the  temperature at a constant  using
the Gaussian constraint thermostat
to eliminate the viscous heating effect.
No difference was
detected  between the profile-based and profile-unbased
thermostats \cite{Evans}, so results with
the profile-based thermostat will be presented in this paper.
Our method of applying shear is as follows:
The system was at rest for $t < 0$ for a very long
equilibration time and was then sheared for  $t > 0$.
Here we added the average velocity $\gdot y$  to the
velocities of all the particles in the $x$ direction
at  $t = 0$
and afterwards maintained  the shear flow by using
the Lee-Edwards boundary condition\cite{Allen,Evans}.
Then steady states  were  realized  after a transient time.
In our case shear flow serves to destroy glassy structures and
produces no long range structure.

\section{Pair correlations and bond breakage}
\setcounter{equation}{0}

\subsection{Pair correlations}
Because of the convenience of visualization in 2D,
 we first present a snapshot of particles
 at $\Gamma_{eff}=1.4$ in 2D in Fig.1,  which
 gives an intuitive picture of the particle
configurations.
We can see that  each particle
is   touching mostly   6  particles and  infrequently
5 particles at distances close to
%%%%%%%% Fig. 1:  2D jammed configuration  %%%%%%%%%%%%%%%%%%%%%%%%%%%%%%%%
$\sigma_{\alpha\beta} (= 1.095^{-1}\ell_{\alpha\beta})$.
Similar jammed particle
configurations can  also  be found
 in 3D, where the coordination number
  of other particles around each particle is  about 12.
Then it is natural that  the
 pair correlation functions $g_{\alpha\beta}(r)$ ($\alpha,
\beta=1,2$)  have a very sharp
peak at $r\cong \sigma_{\alpha\beta}$,
as displayed in Fig.2 for  $\Ge=1.4$ in 2D and $\Ge=1.55$ in 3D.
Furthermore, the heights of these peaks
are  all close to $7$ in 2D and
$4$ in 3D.  This confirms the scaling form
(2.10) around the first peak.
%%%%%%%% Fig. 2: a)g in 2D, b)g in 3D %%%%%%%%%%%%%%%%%%%%%%

We newly introduce  a density variable
 representing  the degree of particle packing  by
\be
\hat{\rho}_{eff}({\bi r})= \sigma_1^d \hat{n}_1({\bi r})
+ \sigma_2^d\hat{n}_2({\bi r}),
\label{eq:3.1}
\en
in terms of which the local volume fraction
of the soft-core regions
is  $\pi \hat{\rho}_{eff}({\bi r})$ in 2D and
by $(4\pi/3) \hat{\rho}_{eff} ({\bi r})$ in 3D.
We also consider
the local composition fluctuation,
\be
\delta\hat{X} ({\bi r})=
\frac{1}{n} [ x_2 \hat{n}_1 ({\bi r}) - x_1 \hat{n}_2 ({\bi r}) ],
\label{eq:3.2}
\en
where $x_1=x_2=1/2$ in our case.
%%%%%%%% Fig. 3: a)S in 2D, b)S in 3D %%%%%%%%%%%%%%%%%%%%%%
In Fig.3 we show the corresponding, dimensionless
 structure factors,
\be
S_{\rho\rho}(q)=  \sigma_1^{-d} \int d{\bi r}
e^{i{\bi q}\cdot{\bi r}}
\av{\delta\hat{\rho}_{eff}({\bi r})
\delta\hat{\rho}_{eff}({\bi 0})},
\label{eq:3.3}
\en
\be
S_{\rho X}(q)=  \sigma_1^{-d} \int d{\bi r}
e^{i{\bi q}\cdot{\bi r}}
\av{\delta\hat{\rho}_{eff}({\bi r})\delta\hat{X} ({\bi 0})},
\label{eq:3.4}
\en
\be
S_{XX}(q)=  \sigma_1^{-d} \int d{\bi r} e^{i{\bi q}\cdot{\bi r}}
\av{\delta\hat{X}({\bi r})\delta\hat{X} ({\bi 0})},
\label{eq:3.5}
\en
where $\delta\hat{\rho}_{eff}=\hat{\rho}_{eff}-
\av{\hat{\rho}_{eff}}$.  They are linear combinations
of the usual structure factors,
\be
S_{\alpha\beta}(q)=  n_{\alpha}n_{\beta}
\int d{\bi r} e^{i{\bi q}\cdot{\bi r}}
[ g_{\alpha\beta}(r) -1],
\label{eq:3.6}
\en
from the definitions (3.1) and (3.2).
The temperatures in Fig.3 are common to those in Fig.2.
Note that the dimensionless wavenumber $q$ is
measured in units of $\sigma_1^{-1}$.
The $S_{\rho\rho}(q)$ has a pronounced
peak at  $q \sim 6 $  and becomes
very small ($\sim 0.01)$ at  smaller $q$
 both in 2D and 3D.
In this sense our systems are  highly
incompressible at long wavelengths.
On the other hand,   $S_{XX}(q)$ has no peak and
is roughly a constant over a very wide  $q$ region, suggesting
no enhancement of the composition fluctuations
and no tendency of phase separation at least in our simulation
times.

>From Fig.3 we may estimate the magnitude of
the isothermal compressibility
$K_{TX}= (\p{n}/\p{p})_{TX}/n$.  In equilibrium
it is expressed in terms of the fluctuation
variances as
\bea
&&k_BT K_{TX} =  n^{-4} \lim_{q \rightarrow 0}
\bigg  [ S_{11}(q)S_{22}(q)-S_{12}(q)^2 \bigg ]
\bigg / S_{XX}(q) \nonumber \\
&&= \frac{\sigma_1^d}{(\sigma_1^d n_1 + \sigma_2^d n_2)^{2}}
\lim_{q \rightarrow 0}
\bigg  [ S_{\rho\rho}(q)- S_{\rho X}(q)^2 \bigg / S_{XX}(q)
\bigg ]
\label{eq:3.7}.
\ena 
The first line was the expression in Ref.\cite{Kirk51},
and the second line follows if use is made of (3.1) and (3.2).
The dimensionless combination $nk_BT K_{TX}$ is equal to
$0.0028$ in 2D and $0.0067$  in 3D.
If we assume that the adiabatic compressibility
$K_{sX}=   (\p{n}/\p{p})_{sX}/n$ is of the same order as
$K_{TX}$, the sound speed $c$ turns out to be  of order $10$
 in units of $\sigma_1/\tau_0$.

Our structure factors were obtained
by time averaging over very long times,
which are $10^5$ for 2D and $10^4$ for 3D.
However, irregular shapes of $S_{XX}(q)$
persisted  at  long wavelengths $q \ls~ 1$.
Such  large scale composition fluctuations
have very long life times ($\gg \tau_{\alpha}$)
and are virtually frozen throughout the simulation.
Therefore, we admit the possibility that
our supercooled  states at low temperatures
 might phase-separate to form  crystalline regions on much
 longer time scales.
On the contrary, the long wavelength
fluctuations of $\hat{\rho}_{eff}$ have much shorter time scales,
probably they vary on  acoustic time scales $\sim 1/cq$.

As is well known, the temperature
dependence of the static pair correlation functions
is  much  milder than that of  the dynamical quantities.
Similarly, their shear dependence is also  mild
 even for $\gdot \tau_{\alpha} \gg 1$  as long as $\gdot \ll 1$. 
In particular, their spatially
anisotropic part is at most a few   percents of their  isotropic part
around  the first peak positions $r \cong \sigma_{\alpha\beta}$ in our case.
This is consistent with the fact that
the attained shear stress in our simulations are
at most a few  percents of the particularly
high pressure  $p$ of our systems.
Note that  the average shear stress $\sigma_{xy}$
in sheared steady states may be related to  the
steady state pair correlation functions
$g_{\alpha\beta}({\bi r})$ as \cite{Hansen}
\be
\sigma_{xy}=
-\frac{1}{2}\sum_{\alpha,\beta}n_{\alpha}n_{\beta}
\int d{\bi r}v_{\alpha\beta}'(r)
\frac{r_{x}r_{y}}{r}
 g_{\alpha\beta}({\bi r}) ,
\label{eq:3.8}
\en
where $r_x$ and $r_y$ are the $x$ and $y$ components of the
vector ${\bi r}$ connecting  particle pairs.
The dominant contribution here arises from  the  anisotropy
at $r \cong \sigma_{\alpha\beta}$.

\subsection{Bond breakage}

Because of the sharpness of the first peak of
$g_{\alpha\beta}({\bi r})$ in our systems,
 we can unambiguously define {\it bonds}
between particle pairs
at distances close to $\sigma_{\alpha\beta}$ in the absence and 
presence of shear.
Such  bonds  will be broken on the
structural ($\alpha$) relaxation time,
because the bond breakage takes
 place on  local configurational
rearrangements.
We define the  bonds as follows.
For each atomic configuration given at time $t_{0}$,
a pair of particles  $i$ and $j$ is considered to be bonded if
\be
r_{ij}(t_{0})= |{\bf r}_{i}(t_{0})-{\bf r}_{j}(t_{0})|\leq
A_1 \sigma_{\alpha\beta},
\label{eq:3.9}
\en
where $i$ and $j$ belong to the species $\alpha$ and $\beta$,
respectively.
We have set $A_1= 1.1$ for 2D and $1.5$ for 3D.
The resultant bond numbers between $\alpha$ and $\beta$
pairs, $N_{b\alpha\beta}$, are related to the first peak structure of
$g_{\alpha\beta}(r)$ as follows. We consider the
coordination  number $\nu_{\alpha\beta}$ of $\beta$ particles
around a $\alpha$ particle
within the distance $A_1\sigma_{\alpha\beta}$ \cite{Bernu},
\be
\nu_{\alpha\beta}= n_{\beta} \int_{r < A_1\sigma_{\alpha\beta}}
d{\bi r} g_{\alpha\beta}(r) \sim
 C n_{\beta}\sigma_{\alpha\beta}^d,
\label{eq:3.10}
 \en
where $C$  is about 5 in 2D and 12 in 3D.
Then we simply have
\be
N_{b\alpha\alpha}= \frac{1}{2} N_{\alpha} \nu_{\alpha\alpha}
\quad ({\alpha=1,2}), \quad
N_{b12}= \frac{1}{2} N_{1} \nu_{12}+ \frac{1}{2} N_2 \nu_{21} .
\label{eq:3.11}
 \en
In 2D at $\Ge=1.4$,  we  find
$\nu_{11}= 2.19$, $\nu_{12}=\nu_{21}=2.54$, and $\nu_{22}=3.41$,
which are consistent with
the bond numbers, $N_{b11}=5514$,
$N_{b11}=13135$, and $N_{b22}=8436$, counted
 in a simulation.
In 3D at $\Ge=1.55$,  these numbers are
$\nu_{11}= 5.57$, $\nu_{12}=\nu_{21}=6.90$, $\nu_{22}=8.30$,
which are again consistent with
$N_{b11}=13925$,  $N_{b11}=34476$, and $N_{b22}=20744$
 in a simulation.
We stress that our bond definition is
insensitive to $A_1$,  owing to the sharpness of the first peak, 
as long as it is somewhat larger than 1 and smaller
than the second peak distances
divided by  $\sigma_{\alpha\beta}$.

After a lapse of time  $\Delta t$,  a pair is regarded
to have been broken if
\be
 r_{ij}(t_{0}+\Delta t) >A_2\sigma_{\alpha\beta}
\label{eq:3.12}
\en
with $A_2=1.6$ for 2D and $1.5$ for 3D.
 This definition of bond breakage is also
 insensitive to $A_2$ as long as  $A_2 \geq A_1$
 and $A_2\sigma_{\alpha\beta}$ is shorter than the second
 peak position of $g_{\alpha\beta}(r)$.
We have followed  the relaxation  of
the total surviving (unbroken) bonds
$N_{bond}(\Delta t)$ from the initial
number
\be
N_{bond}(0)= N_{b11}+N_{b12}+N_{b22}
\label{eq:3.13}
\en
to zero with increasing $\Delta t$.
No significant difference has been found  between
the bond breakage processes  of the three
kinds of bonds, 1-1, 1-2, and 2-2,
so we consider their sum only.
We  define the bond breakage
time $\tau_b$ by
\be
N_{bond}(\tau_b)= N_{bond}(0)/e .
\label{eq:3.14}
\en
The relaxation is not simply  exponential
at low temperatures, apparently
because of large scale
heterogeneities composed of  relatively
weakly and strongly bonded regions.
If we fit  $N_{bond}(\Delta t)$
 to the stretched exponential form,
$N_{bond}(\Delta t) \sim
 \exp [-({\Delta t}/{\tau_b} )^{a'}  ]$,
 the exponent $a'$ is close to 1 at
relatively high temperatures
but is considerably smaller than 1
at the lowest  temperatures (for example,
$a' \sim 0.6$ at $\Gamma_{eff}=1.55$ in 3D).

In Fig.4 we show the bond breakage time
$\tau_b= \tau_b(T)$  in the absence
of shear as a function of
the temperature.
It grows strongly with decreasing the temperature.
As will be shown in (6.8) in Sec. III,
the bond breakage time $\tau_b$ is
proportional to the $\alpha$ relaxation time $\tau_\alpha$
obtained from the decay of the self part of the time 
correlation function $F_s(q,t)$ at $q=2\pi$.  
The shear dependence of the bond breakage time
$\tau_b=\tau_b(\gdot)$ is also of great interest.
As shown in Fig.5,  the bond breakage rate $1/\tau_b (\gdot)$
  consists of the thermal breakage rate $1/\tau_b (0)$
strongly dependent on $T$ and a shear-induced
breakage rate proportional to $\gdot$.  It is expressed
in the simplest conceivable form,
\be
1/\tau_b(\gdot) \cong 1/\tau_b(0) + A_b \gdot ,
\label{eq:3.15}
\en
where  $A_b=0.57$ in 2D and $0.80$ in 3D.
In the strong shear condition $\gdot \tau_b(0) > 1$,
jump motions are induced by shear on
the time scale of $1/\gdot$.
We shall see that  the bond breakage
occurs more homogeneously with
 increasing shear.  Therefore, it is natural that,
 when the strain $\gamma=\gdot \Delta t$ reaches 1,
  a large fraction of bonds have been  broken by
shear.

%%%%%%% Fig. 4 \tau_b(0) in 2D and 3D %%%%%%%%%%%%%%%%%%%%%%
%%%%%%% Fig. 5     \tau_b(\gdot)$     %%%%%%%%%%%%%%%%%%%%%%%%%%%%%%%%%%%%%

\section{Heterogeneity in bond breakage}
\setcounter{equation}{0}

Following the bond breakage process we can
visualize the kinetic
heterogeneity   without ambiguity
 and quantitatively characterize the
 heterogeneous patterns.
In Fig.6 we show spatial distributions of broken bonds
in a time interval of $[t_0, t_0+ 0.05\tau_b]$
in 2D, where about 5$\%$
of the initial  bonds defined at  $t=t_0$ have been  broken.
The dots are the
center positions
${\bf R}_{ij}= \frac{1}{2} ( {\bi r}_i(t_0) +
 {\bi r}_j(t_0))$ of the broken pairs  at
 the initial time $t_0$.
%%%%%%%% Fig. 6  broken bond snapshots  %%%%%%%%%%%%%%%%%%%%%%
The broken bonds are seen to form {\it clusters}
with various sizes.   The heterogeneity is marked in the glassy
case (b) with  $\Gamma_{eff}=1.4$ and $\gdot=0$,
whereas it is much weaker for the liquid case (a)
 with $\Gamma_{eff}=1$ and $\gdot=0$.
 The bond breakage time $\tau_b$ is 17 in (a) and $5\times
 10^4$ in (b).   In (c)
 we set  $\gdot=0.25\times 10^{-2}$ and
 $\Gamma_{eff}=1.4$ with  $\tau_b =32 \sim 1/\gdot$.
 The heterogeneity is known to
 become much  suppressed
 by shear,   while  its spatial  anisotropy remains  small.
  Notice that   even in normal liquids
   bond breakage events frequently occur in the form of
   strings involving a few to
  several particles,   obviously
  because of the high density of our system.
 In glassy  states such strings become longer and
  aggregate forming large scale clusters.
 In 3D we also observe string-like jump motions in accord with
 Ref.\cite{Donati} and aggregation of such strings at low
 temperatures.

  In Fig.7 we write the broken bonds
 in two consecutive time intervals,
$[t_0, t_0+ 0.05\tau_b]$ and $[t_0+ 0.05\tau_b, t_0+ 0.1\tau_b]$ at
$\Ge=1.4$ and $\gdot=0$.
The clusters of  broken bonds in the two time
 intervals  mostly overlap or are adjacent to one another.
This demonstrates
 that {\it weakly bonded regions} or  {\it collectively rearranging
 regions}  (CRR) follow  complex space-time evolution
 on the scales  of $\xi$ and
 $\tau_b$.   We do not know its evolution
 laws  but will encounter a dynamical scaling law between
 $\xi$ and $\tau_b$ in  (4.4) below.
%%%%%%%  Fig. 7 consecutive time intervals of broken bonds %%%

We define the
structure factor $S_b(q)$ of the broken bonds as
\begin{equation}
S_b(q)=\frac{1}{N_b}{\bigg \langle} {\bigg |} \sum_{<i,j>}
\exp(i{\bf q}\cdot{\bf R}_{ij})
{\bigg |}^{2} {\bigg \rangle},
\label{eq:4.1}
\end{equation}
where the summation is over the broken pairs,
$N_b$ is the total number of the broken bonds
in a time interval $[t_0, t_0+\Delta t]$,
and the angular average over the direction of the
wave vector has been  taken.
Furthermore, we have  averaged
over 5--50 $S_b(q)$ data calculated from sequential
configurations of broken bonds.
Fig.8 displays  the resultant
$S_b(q)$ after these averaging procedures
on logarithmic scales at several
$\Ge$ without shear.
%%%%%%%% Fig.8   S_b(q) on log scale without shear %%%%%%%%%%%%
The enhancement of $S_b(q)$ at small $q$
arises from   large scale kinetic heterogeneities
 growing with increasing $\Gamma_{eff}$ both in 2D and 3D.
>From   a plot of $1/S_b(q)$ versus $q^2$
in our previous reports \cite{Yamamoto_Onuki1},
we already  found  that  $S_b(q)$ can  be nicely
fitted to  the Ornstein-Zernike (OZ) form :
\begin{equation}
S_b(q)={S_b(0)}/(1+\xi^{2}q^{2}).
\label{eq:4.2}
\end{equation}
The correlation length $\xi$ is determined from this expression.
It grows up to the system length at the lowest temperatures
and  is insensitive to the width of
the time interval $\Delta t$ as long as
it is considerably shorter
than the bond breakage time $\tau_b$ \cite{Yamamoto_Onuki1}.
The agreement of our $S_b(q)$ with the OZ form becomes  more
evident  in the plots of
$S_b(q)/S_b(0)$ versus $q\xi$ in Fig.9, in which
%%%%%%%% Fig. 9   $S_b(q)/S_b(0)$ versus $q\xi$%%%%%%%%%%%%%%%%%%%%%%
all the data collapse onto a single OZ
master curve  both in 2D and 3D.
In particular,  in 3D  the deviations are very  small,
although   $\xi \sim L$ for low $T$ and small $\gdot$
in our case.

We also notice that $S_b(q)$ is
insensitive to the temperature at large $q$, so from
the OZ form (4.2) we find
\be
S_b(0) \sim \xi^2.
\label{eq:4.3}
\end{equation}
 The clusters of
the broken bonds are  thus very  analogous to the
the critical fluctuations in Ising spin systems.
In fact, small scale
heterogeneities with sizes $\ell$ in the region
 $1 \ll \ell \ll \xi$ are insensitive to the
 temperature.
The relation (4.3) is analogous to
the relation, $\chi \propto  \xi^{2-\eta}$,  in Ising
spin systems  between
the magnetic susceptibility $\chi= \lim_{q \rightarrow 0}S(q)$
and the correlation length $\xi$
near the critical point. Here
$S(q)$ is the spin structure factor and
$\eta$ is the Fisher critical exponent ($\ll 1$  in 3D).

Obviously,
 $\xi$ represents the order of the maximum length
of the clusters.  However,  Adam and Gibbs \cite{Adam}
intuitively expected
that the {\it minimum} size of CRR increases
as $\exp ({\rm{const.}}/(T-T_0))$ on lowering $T$ towards
$T_0$. It has  also been discussed
as to  whether or not
 there is an underlying thermodynamic
 phase transition at a nonzero
 temperature $T_0$  in highly supercooled liquids
 \cite{Dasgupta,Ernst,Ghosh}.
>From our data   we cannot detect
 any  divergence of $\xi$ at a nonzero
 temperature, although this is not conclusive
 due to the finite size
 effect arising from $\xi \sim L$.

Furthermore, as in critical dynamics, we
have confirmed   a dynamical scaling
relation  between
 the bond breakage time  $\tau_b$ and
the correlation length $\xi$,
\be
\tau_b \cong  A \xi^{z} ,
\label{eq:4.4}
\en
where $z=4$  in  2D \cite{Perera_private} and $z=2$  in  3D.
The coefficient $A$
is independent of  $\Ge$ and $\gdot$ chosen in our simulations,
as shown in Fig.10 (a) and (b).
%%%%%%%% Fig. 10 tau_b vs \xi  %%%%%%%%%%%%%%%%%%%%%%%%%%%%%%%%
Notice that  the data points at the largest
$\xi$  in Fig.10  are those at zero shear
for each $\Gamma_{eff}$.
At present  we cannot
explain the origin of  these simple numbers for
 $z$.  We may only argue that
$z$ should be larger in 2D than in 3D because of stronger
configurational restrictions in  2D.
It is surprising that (4.4)  holds even in strong
shear $\gdot\tau_{b}(0) \gg 1$, where
the correlation length is independent of $T$
and is  determined by shear as
\be
\xi \sim \gdot^{-1/z}.
\label{eq:4.5}
\en
In Fig.10b for 3D, however,
we notice $\xi > L$ at $\Ge=1.50$ and $1.55$ for weak shear.
At present we cannot assess
influence of  this finite size effect.

In a zeroth order
approximation, therefore,
 the kinetic heterogeneities are
characterized by a single parameter,
$\xi$ or $\tau_b$, owing to  the
small space anisotropy induced by shear in our systems.
The shear rate $\gdot$ is apparently
playing  a role similar to a magnetic field $h$
in Ising spin systems.
Thus, $\gdot$ and $T$  are
two relevant external  parameters
in supercooled liquids, while   $h$ and the
reduced temperature $(T-T_c)/T_c$ are
two relevant scaling fields in Ising systems.

\section{Supercooled liquid rheology}
\setcounter{equation}{0}

We next  examine nonlinear rheology
in our  fluid mixtures in supercooled amorphous states.
We first display in Fig.11
 the shear-dependent viscosity $\eta (\gdot)$ (in units of
 $\epsilon \tau_0/\sigma_1^d$)  versus $\gdot$  in steady
 states at various $\Gamma_{eff}$ in 2D and 3D.
%%%%%%%% Fig. 11 \eta vs. \gdot %%%%%%%%%%%%%%%%%%%%%%%%%%
This rheological  behavior is similar
 to those in the experiments \cite{Simons1,Simons2,Yue}.
The viscosity is much enhanced at large $\Gamma_{eff}$
(low $T$) and at low shear,
but it tends to be independent of $T$  at very  high shear.
Remarkably,  glassy states exhibit large non-Newtonian
behavior even when $\gdot$ is much smaller than the microscopic
frequency $1/\tau_0=1$,  whereas such large
effects are expected to appear
only for $\gdot \sim 1/\tau_0$ in
normal liquids far from the critical point \cite{Evans,Hanley}.

%%%%%%% Fig. 12 \eta vs. \tau_b %%%%%%%%%%%%%%%%%%%%%%%%%%
In Fig.12 we demonstrate that  the
viscosity $\eta(\gdot)= \sigma_{\alpha\beta}/\gdot$
is determined solely by
the bond breakage time $\tau_b(\gdot)$ in (3.15) as
\be
\eta (\gdot) \cong A_{\eta}\tau_b (\gdot) +
\eta_B  ,
\label{eq:5.1}
\en
where $A_{\eta}$ and
$\eta_B$ are $0.34$ and $ 6.25$ in 2D,
and $0.24$ and $2.2$ in 3D, respectively.
Because
the linearity $\eta \propto \tau_b$ is
systematically violated at
small $\tau_b$, the presence of the background viscosity
 $\eta_B$  independent of
$\Gamma_{eff}$ and $\gdot$ may be  concluded.
Note that the effective exponent
$(\gdot/\eta)(d\eta/d\gdot)$
remains about $-0.8$ in Fig.11.
As well as the
kinetic heterogeneities, steady state rheology
is  determined only
by a single parameter, $\tau_b$ or $\xi$.
This suggests that a sheared steady state can be fairly
mapped onto
a quiescent state with a higher temperature
but with the same $\xi$.

Substitution of  (3.15) then yields
\be
 \eta (\gdot)
\cong A_{\eta}   \bigg /{\bigg [}
 \tau_b(0)^{-1}  + A_b \gdot {\bigg ]} +\eta_B,
\label{eq:5.2}
\en
This form coincides with the empirical law (1.1) by Simons
{\it et al.}\cite{Simons1,Simons2}.    Fig.13 shows
that the ratio
$(\eta (\gdot)-\eta_B)/(\eta (0)-\eta_B)$
can be fitted to the universal curve
$1/(1+ A_b x )$ with $x=\gdot\tau_b(0)$ independently of
$\Gamma_{eff}$ both in 2D and 3D.
%%%%%%%%%%  Fig.13    \eta (\gdot)-\eta_B)/(\eta (0)-\eta_B)%%%
In strong shear $\gdot \tau_b(0) \gg 1$,   we have
the temperature-independent
behavior $\eta (\gdot) \cong (A_{\eta}/A_{b}) /\gdot + \eta_B$,
which is evidently seen  in Fig.11.
If the background viscosity is negligible,
a constant limiting stress follows as
\be
\sigma_{xy} \cong \sigma_{lim}= A_{\eta}/A_{b} ,
\label{eq:5.3}
\en
which holds for
\be
1/\tau_{b}(0) \ll \gdot \ll \sigma_{min}/\eta_B \sim 0.1/\tau_0.
\label{eq:5.4}
\en
Here $\sigma_{lim}$  is  $0.59$ in 2D and $0.30$ in 3D
in units of $\epsilon/\sigma_1^d$ and is
typically a few percents of the pressure
in our systems. The upper bound in (5.4) is very large
in usual glass-forming liquids but should be
 attainable  in colloidal systems, while
 the lower bound can be very small with lowering $T$.

We will argue to derive the above behavior intuitively.
Supercooled liquids   behave as solids against
infinitesimal  strain  on time scales
 shorter than $\tau_b(0)$ even if
the temperature   is considerably above the so-called
 glass transition temperature.
Fluid-like behavior  is realized only after the bond breakage
processes. It is  natural that the viscosity
is of order $\tau_{b}(0)$ in the linear  regime.
This is usually justified from the
time correlation function expression for the viscosity
in terms of the $xy$ component of the
stress tensor \cite{Hansen}.
In strong shear,
on the other hand,
the bond breakage occurs on the time scale of $1/\gdot$.
Upon each bond breakage  induced by shear,
the particles involved
release a potential energy $\epsilon_{r}$
 whose maximum is $\epsilon$.
 There should be a distribution of $\epsilon_r$,
 but let us assume $\epsilon_r \sim \epsilon$ for
 simplicity. It is then  instantaneously
 changed  into energies of
random motions (and probably sounds)
supported by the surrounding
particles.  The heat transport
is   rapid  in this dissipative process.
Because of this and also because of
the background thermal motions superposed,
we have not detected clear temperature
inhomogeneities like  {\it hot spots}
around broken bonds in our simulations.
The heat production rate is estimated as
\be
Q \sim n\epsilon /\tau_b (\gdot) \sim n\epsilon \gdot ,
\label{eq:5.5}
\en
where  $n$ is the number density.
Because  $Q$ is related to the viscosity
by  $Q= \sigma_{xy}\gdot= \eta (\gdot)\gdot^2$,
 we obtain
\be
\sigma_{xy} =
 \eta (\gdot) \gdot   \sim n\epsilon ,
\label{eq:5.6}
 \en
  in high shear, so  $\sigma_{lim}  \sim n\epsilon$.
Due to disordered particle configurations, however,
it is natural to consider   a distribution of
the released energy $\epsilon_r$, which will explain
the viscosity behavior at lower shear.
Such a distribution was calculated for a model foam system
in shear flow by Durian \cite{Durian}.

\section{Motion of tagged particles}
\setcounter{equation}{0}

In this section
we  will follow the motion of tagged particles in a
glassy matrix
both in the absence and presence of shear in 3D.
We will present results only in  three dimensions.
We first plot in  Fig.14
the self part of the density time correlation
function for various $\Ge$ in the usual  zero shear condition,
\be
F_s(q,t)= \frac{1}{N_1}{\bigg \langle}
\sum_{j=1}^{N_1}\exp [i{\bi q}\cdot{\Delta{\bi r}_j(t)} ] {\bigg \rangle},
\label{eq:6.1}
\en
where  $q=2\pi$, $\Delta{\bi r}_j(t)={\bi r}_j(t) - {\bi r}_j(0)
$,  and the
summation is taken over all the particles of the species 1.
%%%%%% Fig.14 %%%%%%%
This function is proportional to the (incoherent) scattering
amplitude from labelled  particles.
 As is well known, this function has a plateau at low
temperatures
($\Ge \gs~ 1.45$ in our case),  during which
the particle is trapped in  a cage.
After a long time the cage eventually breaks,
resulting in   diffusion with a very small diffusion constant
$D$.  In this paper we define the
 $\alpha$ relaxation time $\tau_{\alpha}$ such
 that  $F_s(q,\tau_{\alpha})=e^{-1}$ at $q=2\pi$.

We    generalize
the time correlation function (6.1)
in the presence of  shear flow by introducing a new
 displacement vector of the $j$-th particle as
\be
\Delta{\bi r}_j(t)={\bi r}_j(t)- \gdot \int_0^{t}dt'
y_j(t'){\bi e}_x - {\bi r}_j(0),
\label{eq:6.2}
\en
where ${\bi e}_x$ is the unit vector
in the $x$ (flow) direction.  In this displacement,
the contribution from  convective transport  by
the average flow has been subtracted, which can be known
from the
time derivative,
\be
\frac{\p}{\p t}
\Delta{\bi r}_j(t)={\bi v}_j(t)- \gdot y_j(t){\bi e}_x.
\label{eq:6.3}
\en
To get clear understanding of  the meaning of
this subtraction,
let us consider a Brownian particle placed in shear flow
as a simple example.
On  time scales longer than the relaxation time of its
velocity, its position ${\bi r}(t)$  obeys
\be
\frac{\p}{\p t}
{\bi r}(t)= \gdot y(t){\bi e}_x + {\bi f}(t),
\label{eq:6.4}
\en
where ${\bi f}(t)$ is the Gaussian random force
characterized by
$\av{{f}_{\mu}(t){f}_{\nu}(t')}
=2D\delta_{\mu\nu}\delta (t-t')$ ($\mu,\nu=x,y,z$).
Then the modified displacement vector reads
\be
\Delta {\bi r}(t) \equiv
{\bi r}(t)- \gdot \int_0^{t}dt'
y(t'){\bi e}_x - {\bi r}(0)= \int_0^tdt' {\bf f}(t') .
\label{eq:6.5}
\en
Here the convective effect  does not appear
explicitly and the diffusion behavior
follows as
\be
\av{\Delta {\bi r}(t)^2}= 6Dt.
\label{eq:6.6}
\en

On the other hand, in
 the incoherent scattering amplitude,
$\Delta {\bi r}_j(t)$  in (6.1) should be taken as
the net displacement ${\bi r}_j(t) -{\bi r}_j(0)$
 even in shear flow.  If  $q_x \neq 0$, it  strongly
 depends   on the thickness
of the  scattering region in the $y$ (velocity gradient)
 direction due to a
 position-dependent Doppler effect \cite{Onuki_review,Clark}.
Only for $q_x=0$, it is proportional
to $F_s(q,t)$ in the above definition.

Fig.15 shows $F_s(q,t)$ at $q=2\pi$ for various $\gdot$
with a fixed temperature,
 $\Gamma_{eff}=1.5$ or $k_BT/\epsilon=0.267$ in 3D.
%%%%%%%%%%  Fig. 15 %%%%%%%%%%%%%
Comparison of this figure with  Fig.14 suggests that
applying shear is equivalent to
raising the temperature.   Recall that
we have made the same statement in analyzing the
bond structure factor $S_b(q)$ and the nonlinear rheology.
Also we may define the shear dependent
$\alpha$ relaxation time $\tau_{\alpha}=
\tau_{\alpha}(\gdot)$
by
\be
F_s(q,\tau_{\alpha})=e^{-1}.
\label{eq:6.7}
\en
In Fig.16 we recognize  that $\tau_{\alpha}$ is proportional
to the bond life time $\tau_b$ as
\be
\tau_{\alpha}\cong  0.1\tau_{b}.
\label{eq:6.8}
\en
This relation holds for  any $\Ge$ and $\gdot$
in our 3D simulations.
%%%%%%%%%%  Fig.16  %%%%%%%%%%%%%
The decay of
$F_s(q,t)$ is not exponential for large $\tau_{\alpha}$.
If it is fitted to the stretched exponential form
$\exp [- (t/\tau_{\alpha})^a ]$ around $t \sim \tau_{\alpha}$,
 the exponent $a$ is increased  from values about $0.8$ to 1 with
increasing $\gdot$ as well as with raising $T$.
Furthermore,
the time correlation function (6.1) has turned out
to be almost independent
of  the direction of the wave vector
$\bi q$.

Next it is convenient to analyze the mean square
displacement of   tagged  particles of the species 1,
\be
\av{(\Delta{\bi r}(t))^2}= \frac{1}{N_1}
\sum_{j=1}^{N_1} \av{(\Delta{\bi r}_j(t))^2}
\label{eq:6.9}
\en
%%%%%%%%%  Fig.17  Fig. 18  %%%%%%%%%%%%%
Fig.17 shows the transition from the ballistic behavior
$\av{(\Delta{\bi r}(t))^2} \cong 3(k_BT/m_1)t^2$ to
the diffusion behavior $\av{(\Delta{\bi r}(t))^2} \cong 6Dt$
in shear flow at $\Gamma_{eff}=1.55$.
The arrows in the figure indicate the $\alpha$
relaxation time $\tau_{\alpha}(\gdot)$.
The crossover occurs around
 $t \sim  \tau_{\alpha}$.
  Fig.18  demonstrates  the surprising isotropy of the statistical
 distribution of $\Delta{\bi r}_i(t)$, where
the mean square displacements of the $x$, $y$, and $z$ components
of the vector   $\Delta{\bi r}_j(t)$ are separately
displayed.
We can thus determine $D$ from the mean square displacement
in addition to
 $\tau_{\alpha}$ in shear flow.
Note that the $x$ component in Fig.18 is not the
usual mean square displacement due to the second term in (6.2).
In the appendix  we will consider the variances of the
net displacement vector ${\bi r}_j(t)-{\bi r}_j(0)$.

Fig.19  shows the shear rate dependences of the
viscosity
$\eta$ $(\sim  \tau_{\alpha})$
 and the inverse diffusion constant $D^{-1}$
from the linear $(\gdot \ls~ 10^{-5})$
to the non-Newtonian regime at $\Gamma_{eff}=1.55$ in 3D,
where   $D$ is measured in units of $\sigma_1^2/\tau_0$ and 
$\eta$ in units of $\epsilon  \tau_0/\sigma_1^d$.
%%%%%%%%%% Fig.19 %%%%%%%%%%%
We deduce the relation
$D^{-1}\sim\gdot^{-\nu}$ with $\nu=0.75 \sim 0.80$
in agreement of the experiment \cite{Silescu},
which is appreciably milder than the viscosity decrease
$\eta \sim \tau_{\alpha} \sim \gdot^{-1}$.
In Fig.20 we plot
$D$ versus $\eta /k_BT$ (in units of $\tau_0/\sigma_1^d$)
obtained for various
$\Gamma_{eff}$ and $\gdot$.
%%%%%%% Fig.20 %%%%%%%%%%%%%%%
The Einstein-Stokes formula, which
 holds  excellently in normal liquids,
 appears to be   violated in supercooled liquids
 as the other simulations have suggested
  \cite{Mountain,Perera_private}.
It is widely believed  that this breakdown is a natural
consequence of the dynamic heterogeneity in glassy states
\cite{Silescu,Ci95,St94}.  Detailed numerical
analysis will appear  in a forthcoming paper.

In our case $\eta/k_BT$ changes over 4 decades until $\xi$
reaches the system dimension $L$, whereas it has been changed
over 12 decades in the experiments \cite{Silescu,Ci95}.
Though the same tendency indicating the breakdown of the
Einstein-Stokes relation has been obtained in our simulation,
we should admit that
our   system size in 3D is  not yet
 sufficiently large and
 our data at $\Ge=1.5$ and $1.55$ might be
 somewhat affected by the system size effect. It is worth noting that the
Monte Carlo simulation of a dense polymer
by Ray and Binder \cite{Ray} shows that the monomer diffusion
constant decreases with increasing the
 system size.

\section{Summary and discussions}
\setcounter{equation}{0}

Most of our  findings in this work have been obtained
from numerical analysis only without first principle
derivations. Nevertheless, we believe that
 they pose new problems and suggest new experiments.
We make some discussions mentioning  possible
experiments  below.\\
1)
Introducing the concept of bond breakage,
we have succeeded
to quantitatively analyze the kinetic heterogeneities
in simple model systems,
which have been witnessed by a number of the authors.
As shown in Fig.6,  strings composed of
 broken bonds are  very  frequent and they  aggregate
 at low temperatures to  form clusters.
The bond breakage time $\tau_b$ is related to
the correlation length $\xi$ as (4.4).
In  future work
we should  clarify  the relationship of our patterns
in the $\alpha$ relaxation and those by
Muranaka and Hiwatari \cite{Muranaka1}
on a much shorter time scale.
\\
2)
The weakly bonded regions identified by the
bond breakage
are purely dynamical objects.
Large scale heterogeneities have not been
clearly detected  in snapshots of the usual
physical quantities such as
the densities, the stress tensor, the kinetic
energy (=temperature), {\it etc}.    On the other hand,
in granular matters  in shear flow \cite{Behringer},
stress heterogeneities have been observed
optically by using birefringent materials.
We admit the possibility
that  such stress heterogeneities
 also exist in supercooled liquids but are
 masked by the thermal fluctuations.  We  will
 check this point in future.
\\
3)
It is of great interest how the kinetic heterogeneities,
which satisfy the dynamic scaling (4.4),
evolve in space and time
and why they look so similar to
the critical fluctuations in  Ising systems
in the mean field level. In our steady state
problem  $T$ and $\gdot$
are two relevant scaling fields, the {\it critical
point} being located at $T=\gdot=0$. No divergence
 has been detected at a nonzero
temperature in our simulations. \\
4)
In his  experiments
Fischer \cite{Fischer} has reported
large excess light scattering
with a correlation length $\xi$ ($20 \sim 200$ nm)
which increases on  approaching
the glass transition from a liquid state.
This indicates the presence of very
large scale {\it density} heterogeneities in
supercooled  liquids,
which is often called Fischer's clusters.
Motivated by  this effect,
Weber {\it et al}\cite{Weber} performed Monte Carlo
simulations on a dense polymer and found
that short range nematic orientational order can
give rise to  enhancement of long range density fluctuations.
They expected  that such anisotropic interactions could
be  the origin of Fischer's clusters.
This suggests that  Fischer's clusters
 do not  exist in liquids composed of
structureless particles.
\\
5)
We have examined  nonlinear rheology in glassy states.
The rheological  relations  obtained are
simplest among those consistent with
the experiments \cite{Simons1,Simons2,Yue}.
 The mechanism of the non-Newtonian
behavior in supercooled liquids
is conceptually new and should be further
examined in experiments such as in colloidal systems
in glassy states.
In particular, polymers should exhibit
pronounced non-Newtonian behavior,
as the glass transition is approached,
even without entanglement.
Rheology of chain systems remains totally unexplored near the
glass transition.\\
6)
In our systems small anisotropic changes of the pair
correlation functions $g_{\alpha\beta}(r)$ near the first peak
($\sim \sigma_{\alpha\beta}$) can give rise to
the limiting shear stress $\sigma_{lim}$,
which is  $3\sim 5\%$ of the pressure in our case.
Note that our systems are highly compressed
with high pressure.
However, the pressure needs not be very high
in supercooled liquids  in the presence of an attractive part
of the potential.
Even in such cases, we expect  that $\sigma_{lim}$
is   a few $\%$ of the shear modulus.
This is  suggested by  the previous
 work on  amorphous alloys
\cite{HSChen,Spaepen,Aragon,Takeuchi,Takeuchi_review},
where the yield stress $\sigma_y$ in  the
  inhomogeneous case (in which shear bands appear)
 is known to be $2 \sim 3 \%$ of the shear modulus. \\
7)
 Stillinger expected that in fragile glass-forming liquids
 shear flow occurs  by {\it tear and repair}  of
 slipping walls  separating   strongly bonded  regions
  \cite{Stillinger}. We have not observed
  such  localization of slips or jump  at least in   our temperature range.
 But there might be a tendency that broken bonds form surfaces
at low temperatures in 3D,
though not conspicuous, which should be checked in
 future.  \\
8) There is no tendency of
phase separation for the parameters used.
However, there are
 many cases in which
the composition fluctuations are enhanced towards
the glass transition temperature.
It is of great interest how  the two transitions
influence each other \cite{Jackel_p,Max}.
 It is also known that shear flow can
 induce  composition fluctuation
enhancement  in asymmetric viscoelastic mixtures,  when
emergence of less viscous regions
can reduce the effective viscosity
\cite{Onuki_review}.
We expect that
this effect can come into play also in supercooled
liquids, for example, for large enough size ratios or in the
presence of  small attraction between the two components.
Experiments to detect this effect seem to be promising in
colloidal systems. \\
9)
 We have introduced the time correlation function $F_s(q,t)$
in shear and found its simple relaxation behavior in
Fig.15. It  coincides with the usual time
correlation function for $q_x =0$ or when
the scattering vector is perpendicular to the flow direction.
Dynamic scattering experiments in shear flow
would be very informative
to detect the shear-induced  diffusion \cite{Onuki_review}.
A direct
diffusion measurement in sheared supercooled fluids
is also very interesting,  which we will be analyzed  in the appendix.
Though our system size is still too small,
we have  detected  a tendency of  the breakdown of
the Einstein-Stokes relation in 3D to obtain
$D \sim \eta^{-\nu}$ with $\nu =0.75 \sim 0.8$. \\
10)
In strong shear
 the structural relaxation is  characterized
 by   $\tau_{\alpha} \sim 0.1\tau_b \sim 0.1/\gdot$
 as (6.8).  This nonlinear effect could be measured
 as drastic reduction of the
 rotational relaxation  time by
 dielectric response or  by more sophisicated techniques \cite{Silescu,Ci95}
 from sheared supercooled liquids. The  same effect
 is expected  for periodic shear flow. \\
11)
Understanding of transient mechanical response
in terms of the kinetic heterogeneities
is of great importance. For example, we have found a
stress overshoot after application of shear strain
in accord with the experiments \cite{Simons1,Simons2}.
We should also understand  glassy behavior of
the complex shear modulus against
small periodic shear \cite{Nagel}.
On these topics we will report shortly.
\\
12)
In our systems we have not yet
found  essential  differences
between  2D and 3D except for
the difference in the value of the dynamic exponent $z$ in (4.4).
We believe that a large part of
essential ingredients of glassy dynamics
can be understood even in two dimensions.
\\
13)
In a forthcoming paper  we
will focus our attention on
jump motions of particles over distances longer than
$\sigma_1$.
They will be shown to occur heterogeneously in space
and determine the diffuison constant.
These heterogeneity  structures are essentially the same
 as those in the bond breakage processes
  studied in this paper.
\\

\acknowledgments
%\section*{ACKNOWLEDGMENTS}

We thank Dr. T. Muranaka, Professor Y. Hiwatari,
Professor K. Kawasaki,
Professor P. Harrowell,
  Dr. D. Perera, and Professor A.J. Liu
  for helpful discussions. Thanks are also due to
  Professor S. Takeuchi, who kindly sent his work on
  amorphous alloys.
This work is supported by Grants in Aid for Scientific
Research from the Ministry of Education, Science and Culture.
Calculations have been carried out at the Supercomputer
Laboratory, Institute for Chemical Research, Kyoto University
and the Computer Center of the Institute for Molecular Science,
Okazaki, Japan.

%\appendix
\section*{Appendix}
%{\bf APPENDIX}
\setcounter{equation}{0}
\renewcommand{\theequation}{A.\arabic{equation}}

Let us calculate the variances
among the $x$, $y$, and $z$ components,
$x_j (t) -x_j(0)$, $y_j(t)-y_j(0)$,  and $z_j(t)-z_j(0)$,
of the net displacement
vector ${\bi r}_j(t) -{\bi r}_j(0)$
of the $j$ th particle in shear flow.
We fix its initial position ${\bi r}_j(0)$  at  ${\bi r}_0
=(x_0,y_0,z_0)$.
The average displacement arises from  the convection as
\be
\av{{\bi r}_j (t) -{\bi r}_j(0)}=\gdot t y_0 {\bi e}_x.
\label{eq:A.1}
\en
Assuming the isotropy of the subtracted displacement (6.2),
which is suggested by  Fig.18,
we may write the variances of the $y$ and $z$
components are
\be
G(t) =\av{(y_j(t)-y_j(0))^2}=\av{(z_j(t)-z_j(0))^2}  .
\label{eq:A.2}
\en
The variance of the $x$ component then becomes
\be
\av{(x_j(t)-x_j(0)- \gdot t y_0)^2}=
G(t) + 2\gdot^2 \int_0^t dt_1 (t-t_1) G(t_1) .
\label{eq:A.3}
\en
The cross correlation exists between  the $x$
and $y$ components as
\be
\av{(x_j(t)-x_j(0))(y_j(t)-y_j(0))}=
\gdot \int_0^t dt_1G(t_1).
\label{eq:A.4}
\en
In the diffusion time regime $t \gs~ \tau_{\alpha}$  we may set
$G(t)=2Dt$ to obtain
\be
\av{(x_j(t)-x_j(0)- \gdot t y_0)^2} \cong
2Dt ( 1+ \frac{1}{3} \gdot^2 t^2 ),
\label{eq:A.5}
\en
\be
\av{(x_j(t)-x_j(0))(y_j(t)-y_j(0))} \cong
D \gdot t^2.
\label{eq:A.6}
\en
Note that $D$ is strongly dependent on $\gdot$ in strong shear as shown
in Fig.19. Measurements
of the above variances are very informative.

%%%%%%%%%%%%%%%%%%%% Figures  %%%%%%%%%%%%%%%%%%%%%%%%%%%%%%%%%%%%

\clearpage

%%% Fig.1 %%%%%%
\begin{figure}
\vspace*{60mm}
\epsfxsize=3.0in
\centerline{\epsfbox{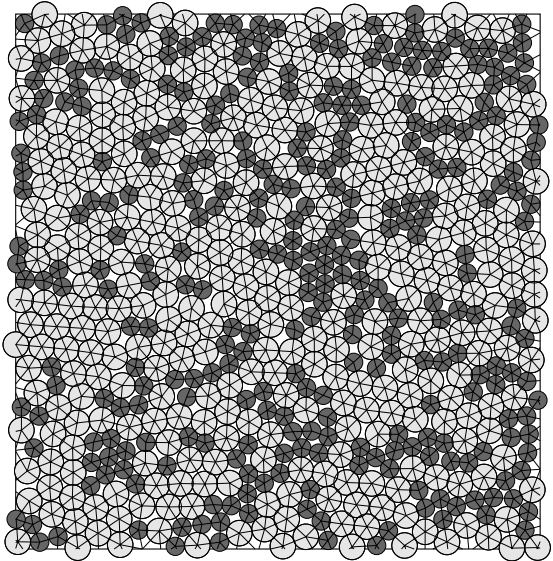}}
\caption{\protect\narrowtext
A typical particle configuration and the bonds defined
at a given time at $\Gamma_{eff}=1.4$ in 2D.
The diameters of the circles here are equal to $\sigma_{\alpha}$.
The areal  fraction of the soft-core regions is $93\%$.
A $1/16$ of the total system is shown.
}
\vspace{70mm}
\end{figure}

%%%%% Fig. 2 %%%%%%%%%%%
\begin{figure}[p]
\vspace*{10mm}
\begin{center}
\epsfxsize=3.0in
\epsfbox{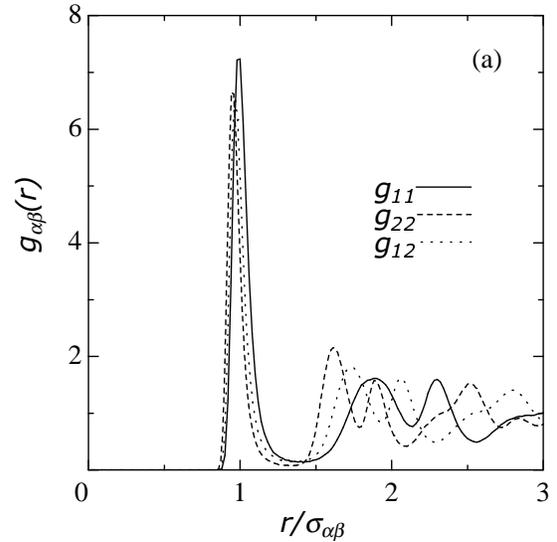}
\vspace{15mm}
\epsfxsize=3.0in
\epsfbox{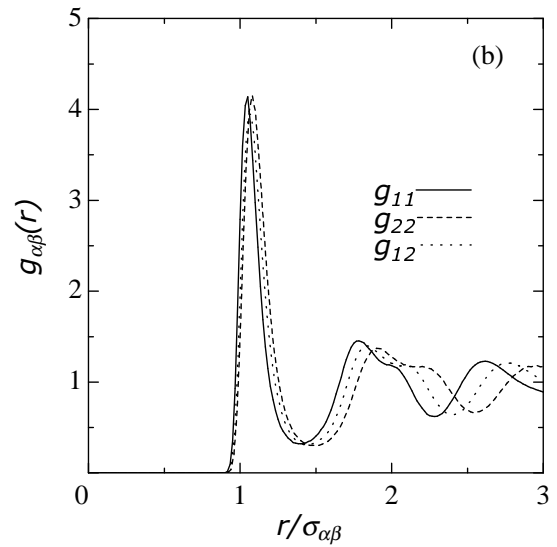}
\vspace{15mm}
\end{center}
\caption{\protect\narrowtext
The pair correlation functions $g_{\alpha\beta}(r)$
in quiescent states  as functions of $r/\sigma_{\alpha\beta}$
at $\Gamma_{eff}=1.4$ in 2D (a)
and at $\Gamma_{eff}=1.55$ in 3D (b).
}
\vspace{70mm}
\end{figure}

%%%%% Fig. 3 %%%%%%%%%%%
\begin{figure}[p]
\vspace*{10mm}
\begin{center}
\epsfxsize=3.0in
\epsfbox{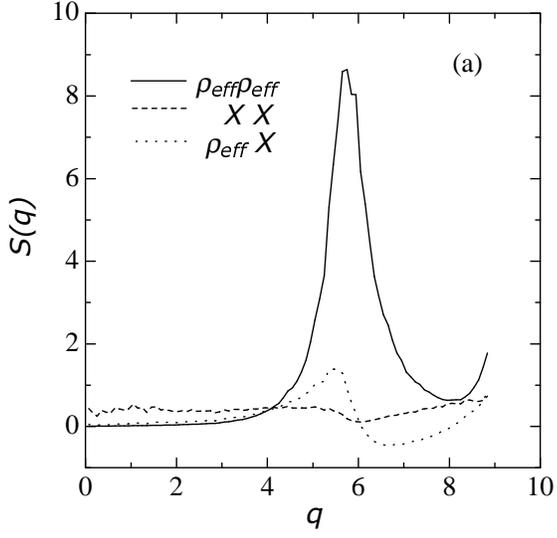}
\vspace{15mm}
\epsfxsize=3.0in
\epsfbox{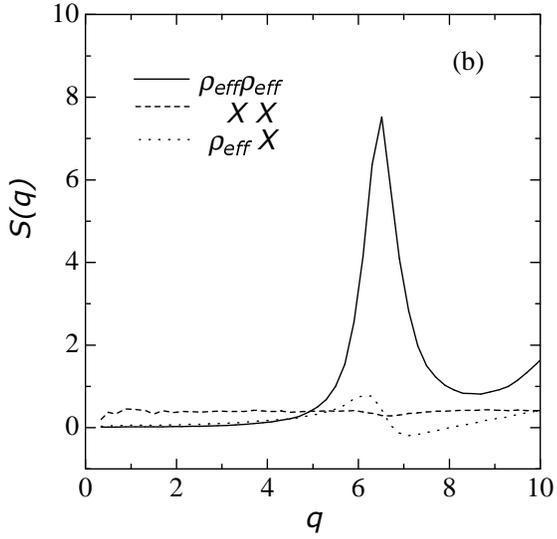}
\vspace{15mm}
\end{center}
\caption{\protect\narrowtext
The structure factors $S(q)$ defined in (3.3)- (3.5)
in quiescent states  at $\Gamma_{eff}=1.4$ in 2D (a)
and at $\Gamma_{eff}=1.55$ in 3D (b).
The dimensionless wavenumber $q$ is
measured in units of $\sigma_1^{-1}$.
The solid, dashed, and dotted lines correspond to $\rho_{eff}-\rho_{eff}$,
$X-X$, and $\rho_{eff}-X$ correlations,
respectively.
}
\vspace{70mm}
\end{figure}

%%%%%%%%%%%Fig. 4 %%%%%%%%%%%%%%%
\begin{figure}
\vspace*{50mm}
\begin{center}
\epsfxsize=3.0in
\epsfbox{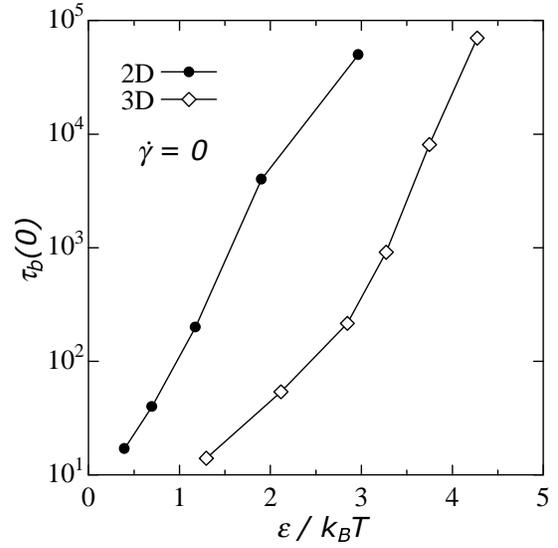}\vspace{-5mm}
\end{center}
\caption{\protect\narrowtext
Temperature dependence of the bond breakage time
$\tau_{b}(0)$ at zero shear ($\bullet$) in 2D
and ($\diamond$) in 3D. The $\epsilon$ is the potential
parameter in the soft-core potentials (2.1).
The time  is measured in
 units of $\tau_0$ in (2.2), so
 $\tau_{b}(0)$ is dimensionless. }
\vspace{50mm}
\end{figure}

%%%%%% Fig.5 %%%%%%%%%%
\begin{figure}
\vspace*{10mm}
\begin{center}
\epsfxsize=3.0in
\epsfbox{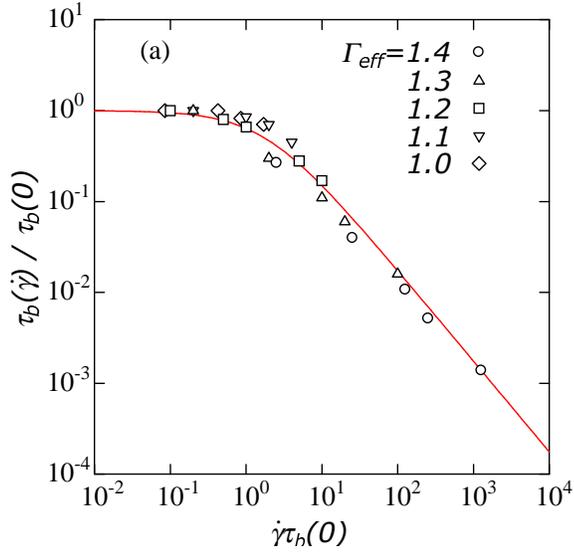}
\vspace{15mm}
\epsfxsize=3.0in
\epsfbox{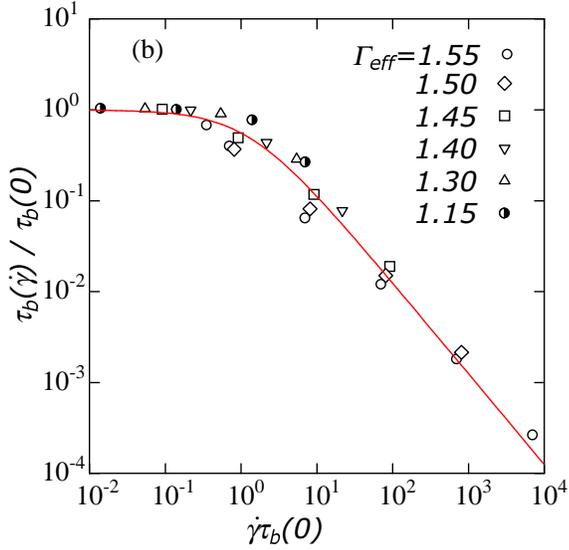}
\vspace{15mm}
\end{center}
\caption{\protect\narrowtext
The normalized  bond breakage time
$\tau_b(\gdot)/\tau_b(0)$ versus
$\gdot \tau_b(0)$ for  various
$\Ge$ in 2D (a) and 3D (b). All the data collapse on
the curve $1/(1+A_bx)$ with $x= \gdot \tau_b(0)$.
}
\vspace{5mm}
\end{figure}

%%%%% Fig. 6 %%%%%%%%%%%
\begin{figure}
\begin{center}
\epsfxsize=2.4in
\hspace*{5mm}\epsfbox{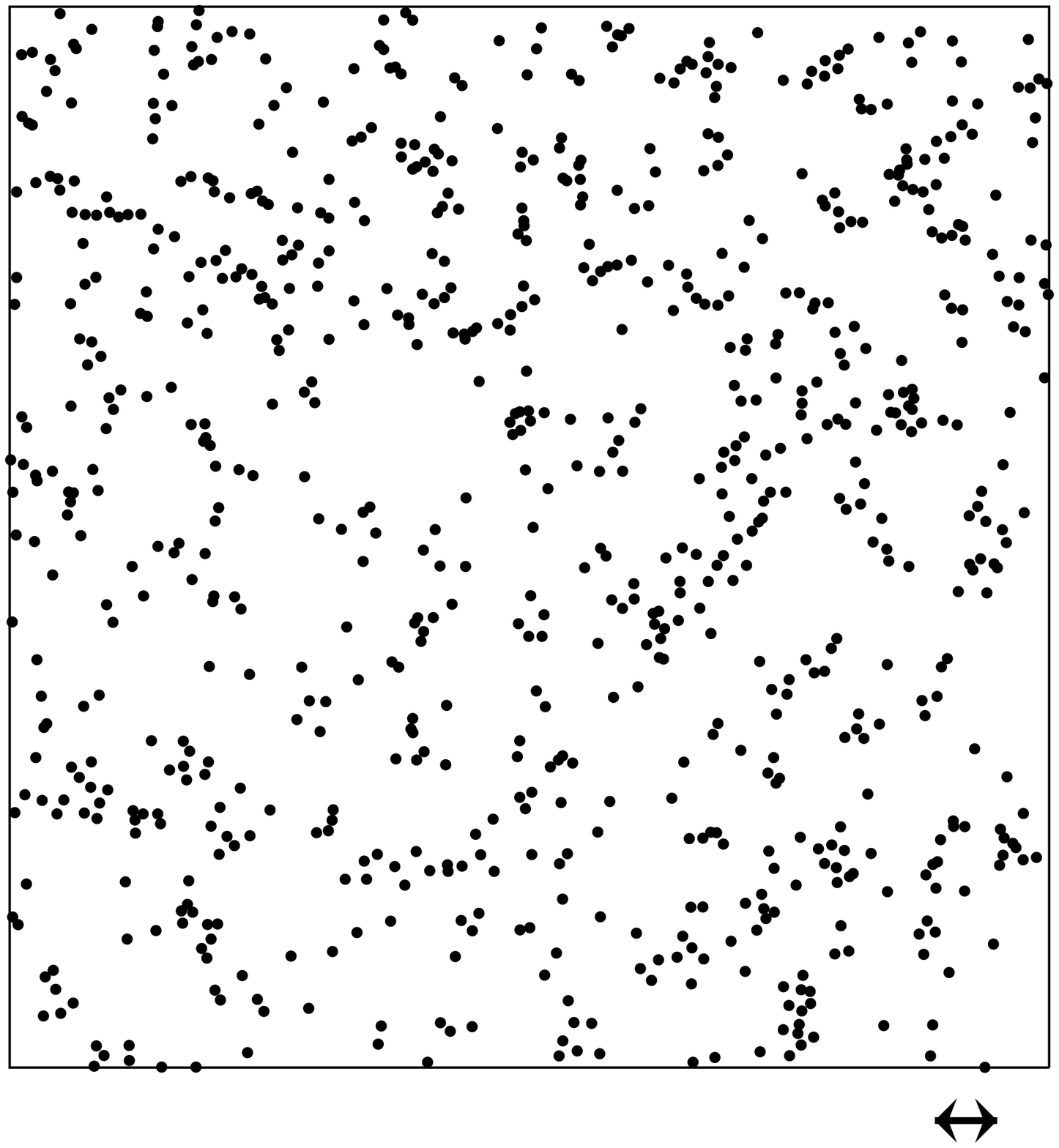}\vspace{-2mm}\\
{\small (a) $\Gamma_{eff}=1$, $\dot{\gamma}=0$}\vspace{3mm}\\
\epsfxsize=2.4in
\hspace*{5mm}\epsfbox{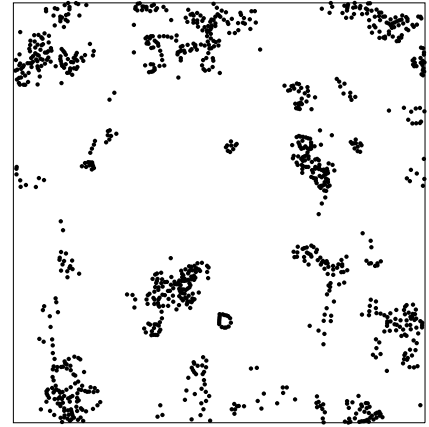}\vspace{-2mm}\\
{\small (b) $\Gamma_{eff}=1.4$, $\dot{\gamma}=0$}\vspace{3mm}\\
\epsfxsize=2.4in
\hspace*{5mm}\epsfbox{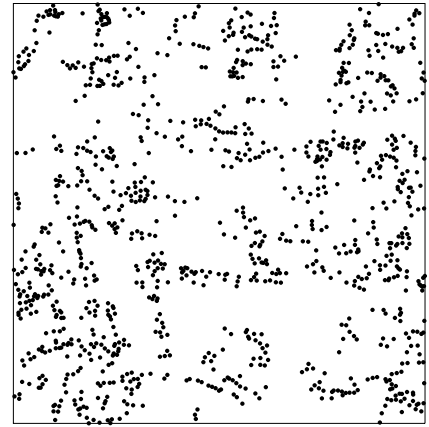}\vspace{-2mm}\\
{\small (c) $\Gamma_{eff}=1.4$, $\dot{\gamma}=0.25\times 10^{-2}$}\vspace{-5mm}
\end{center}
\caption{\protect\narrowtext
Snapshots of the broken bonds in 2D without shear.
The system length is $118 \sigma_1$.
Here $\Ge=1$ with weak heterogeneity  (a),  and
$\Ge=1.4$ with enhanced heterogeneity  (b).
For $\gdot =2.5\times 10^{-2}$
(c), the heterogeneity
is much suppressed.
The flow is in the upward direction and
the velocity gradient is in the
horizontal direction from
left to  right. The arrows indicate the correlation
length $\xi$ obtained from (4.2).
}
\end{figure}
%%%%%%%%%%% Fig.7 %%%%%%%%%%%%%%%
\begin{figure}[p]
\vspace*{45mm}
\begin{center}
\epsfxsize=3.0in
\epsfbox{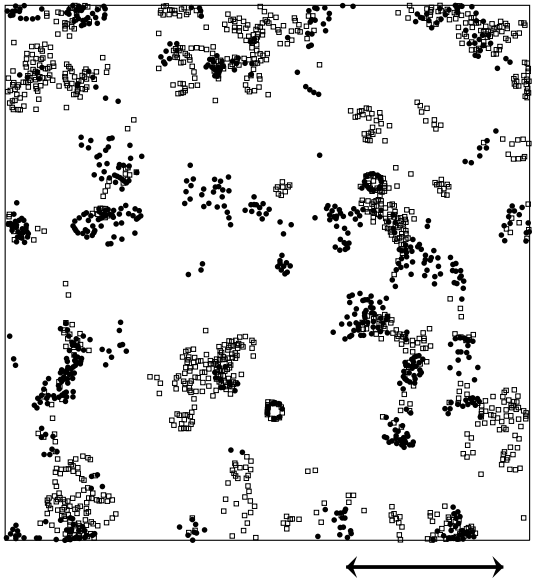}\vspace{2mm}
\end{center}
\caption{\protect\narrowtext
Broken bond distributions in  two consecutive  time intervals,
 $[t_0,t_0+ 0.05\tau_b]$ ($\Box$) and
 $[t_0+0.05\tau_b,t_0+ 0.1\tau_b]$ ($\bullet$),
at $\Gamma_{eff}=1.4$ in 2D. The arrow indicates
$\xi$.
}
\vspace{80mm}
\end{figure}

%%%%% Fig. 8 %%%%%%%%%%%
\begin{figure}[p]
\vspace*{10mm}
\begin{center}
\epsfxsize=3.0in
\epsfbox{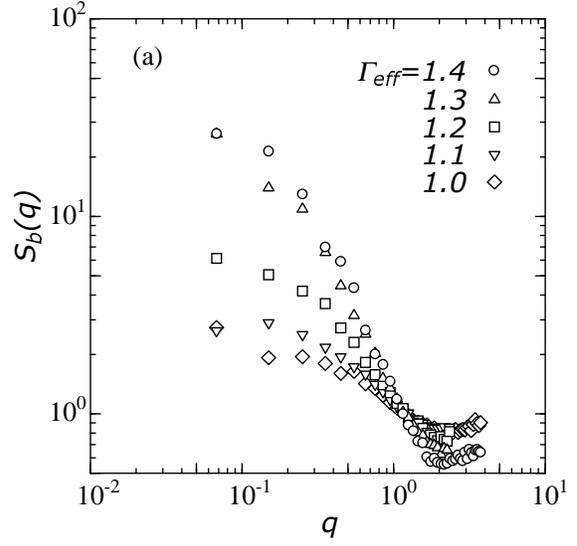}
\vspace{15mm}
\epsfxsize=3.0in
\epsfbox{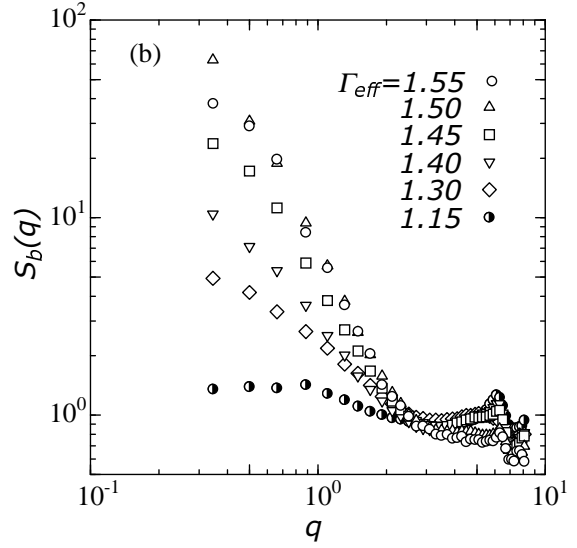}
\vspace{15mm}
\end{center}
\caption{\protect\narrowtext
$S_b(q)$ versus $q$ on logarithmic scales for various
$\Gamma_{eff}$ at $\gdot=0$  in 2D (a) and 3D (b).
Its long wavelength limit is of order  $\xi^{2}$ as (4.3).
}
\vspace{10mm}
\end{figure}

%%%%% Fig. 9 %%%%%%%%%%%
\begin{figure}[p]
\vspace*{10mm}
\begin{center}
\epsfxsize=3.0in
\epsfbox{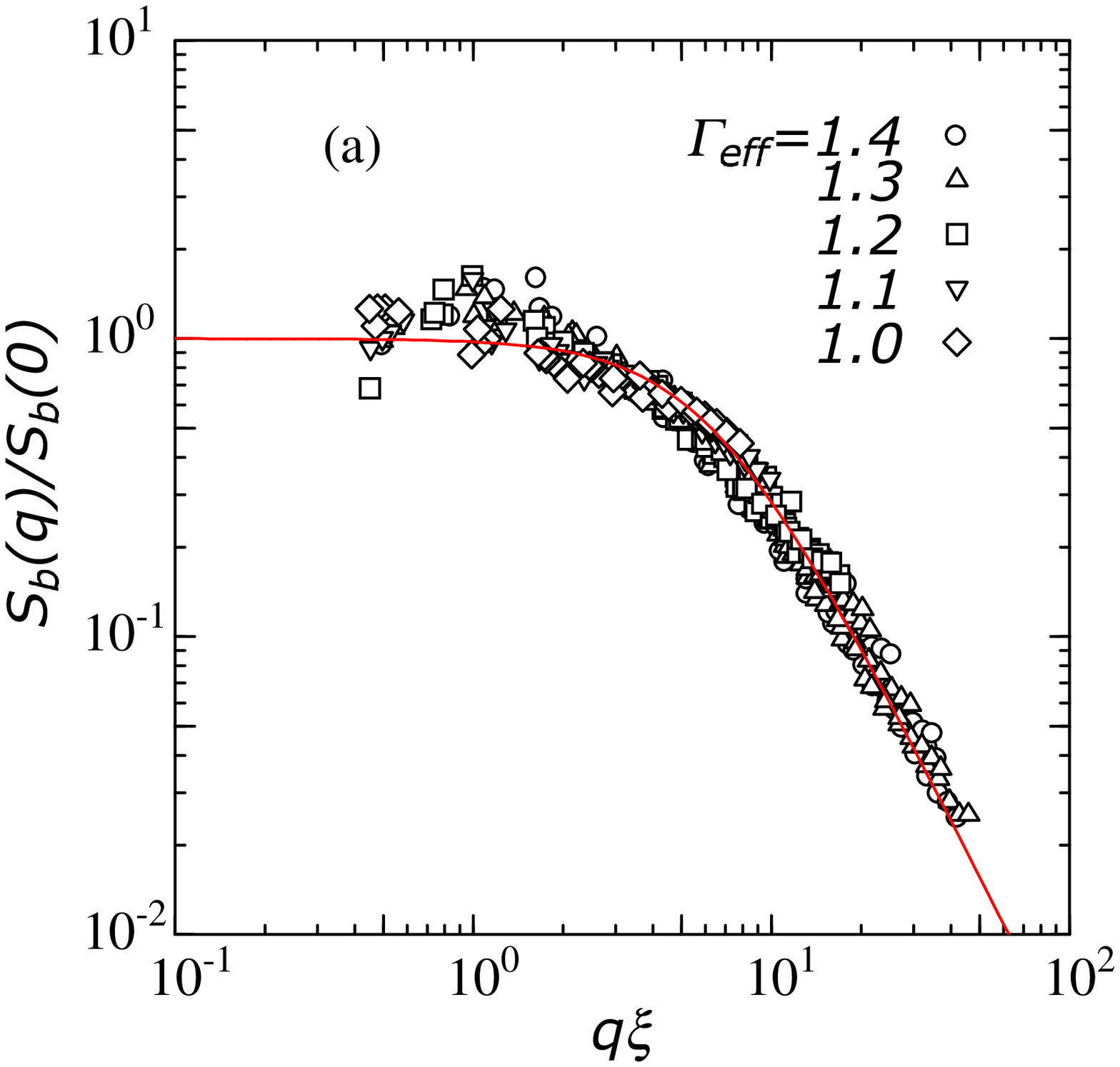}
\vspace{15mm}
\epsfxsize=3.0in
\epsfbox{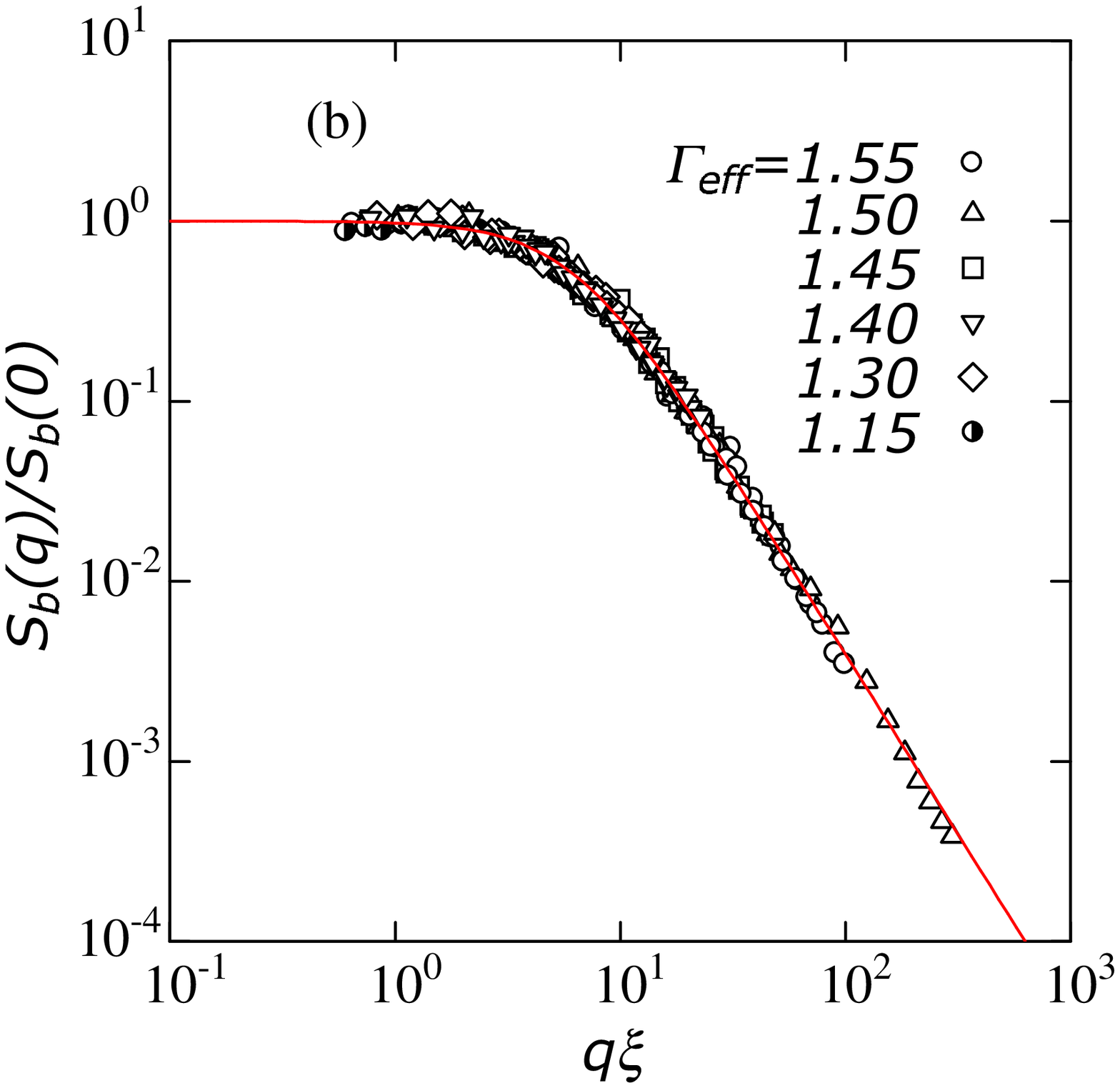}
\vspace{15mm}
\end{center}
\caption{\protect\narrowtext
$S_b(q)/S_b(0)$ on logarithmic scales for various
$\Gamma_{eff}$ and $\dot{\gamma}$ in 2D (a) and 3D (b).
The solid line is the Ornstein-Zernike form $1/(1+x^2)$
 with $x=q\xi$.
}
\vspace{10mm}
\end{figure}

%%%%% Fig. 10 %%%%%%%
\begin{figure}[p]
\vspace*{10mm}
\begin{center}
\epsfxsize=3.0in
\epsfbox{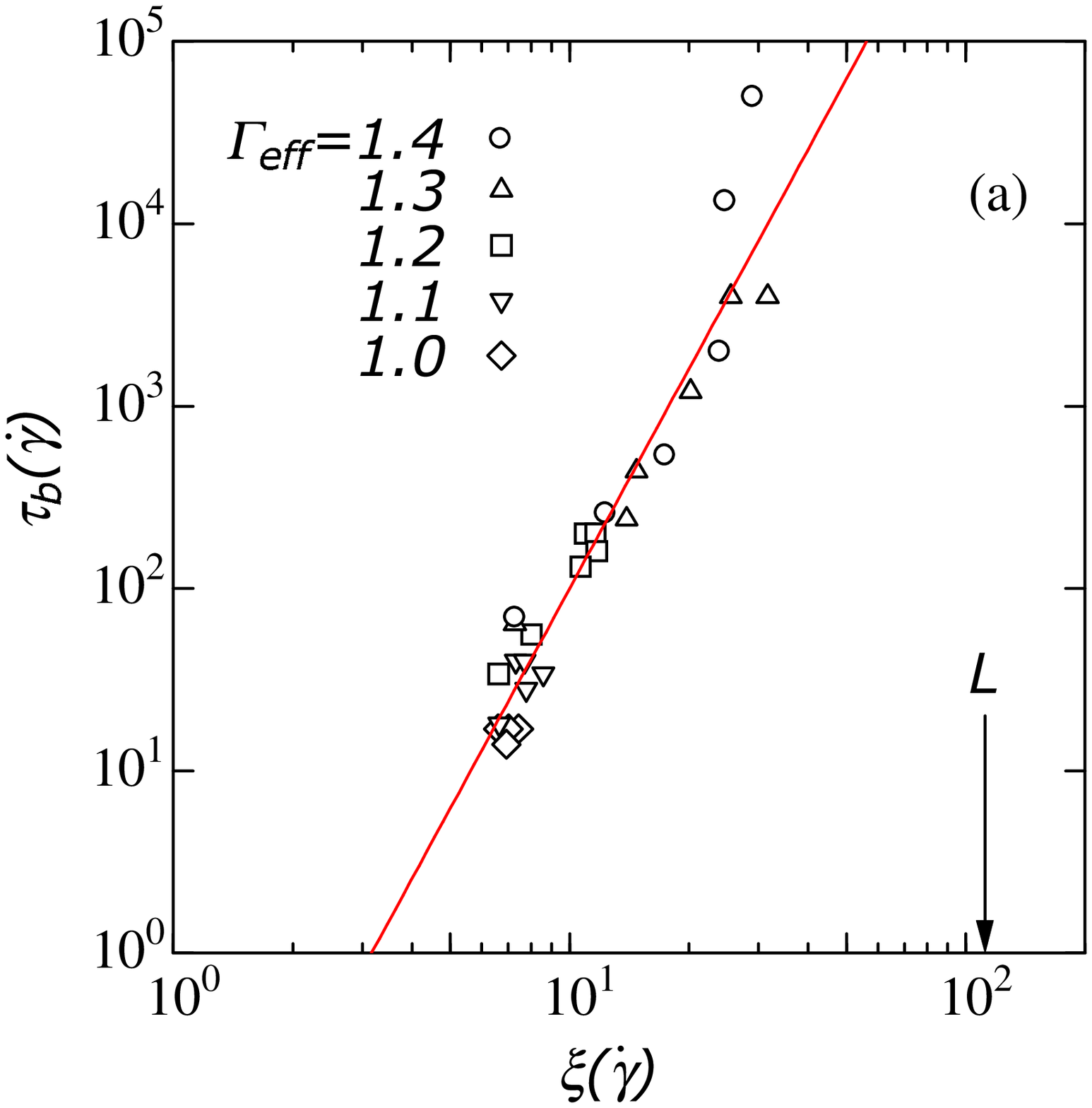}
\vspace{15mm}
\epsfxsize=3.0in
\epsfbox{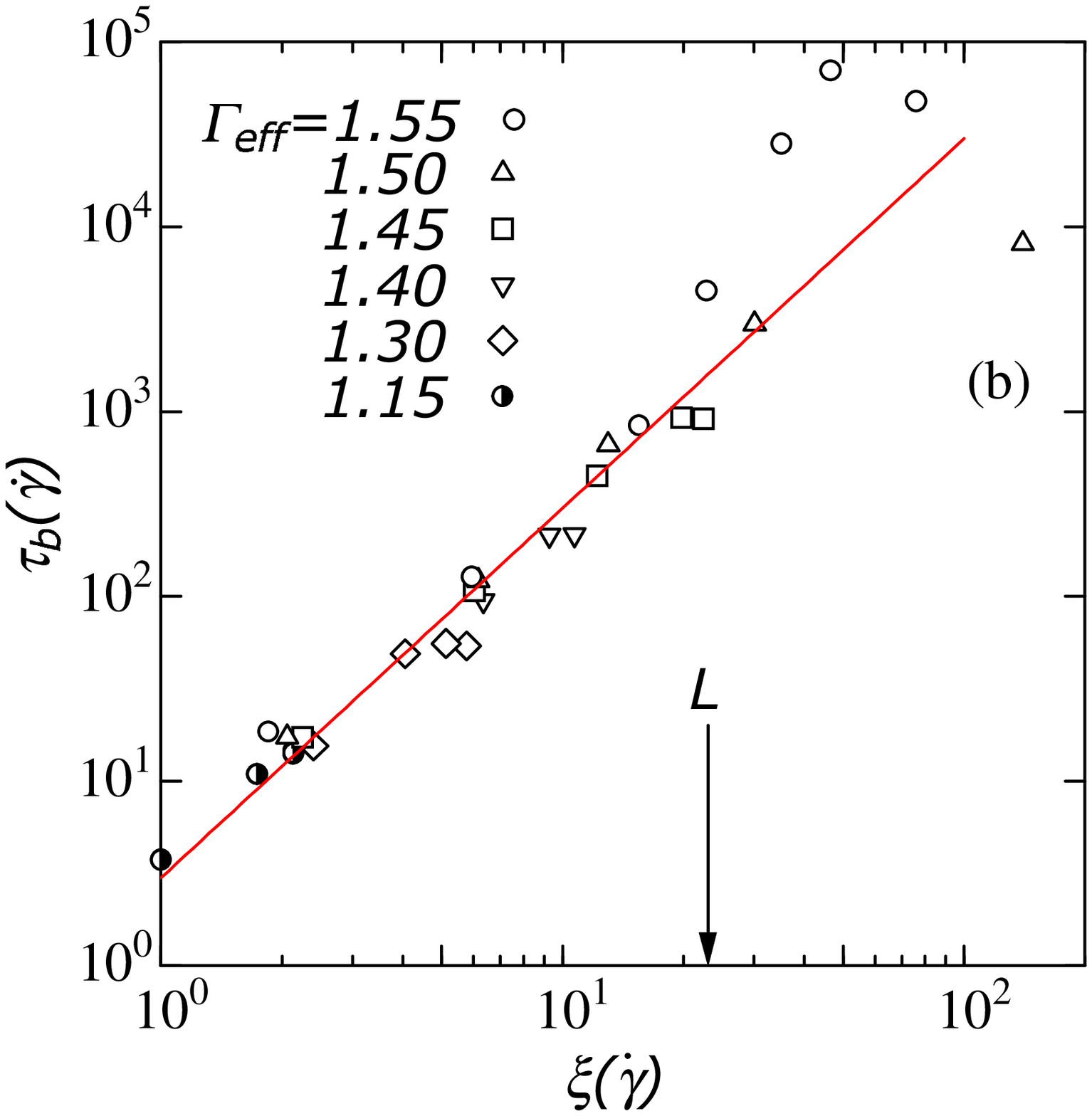}
\vspace{15mm}
\end{center}
\caption{\protect\narrowtext
Universal relation between  the correlation length
$\xi (\gdot)$ in units of $\sigma_1$
and the bond breakage  time $\tau_b (\gdot)$
in units of $\tau_0$ in (2.2).  In 2D (a), the
line of the slope $4$ is a viewing guide
and $L$ is the system length.
The corresponding 3D plot is shown in (b) with
 the slope being $2$.}
\vspace{10mm}
\end{figure}

%%%%%%%%%  Fig. 11 %%%%%%%%%%%%
\begin{figure}[p]
\vspace*{10mm}
\begin{center}
\epsfxsize=3.0in
\epsfbox{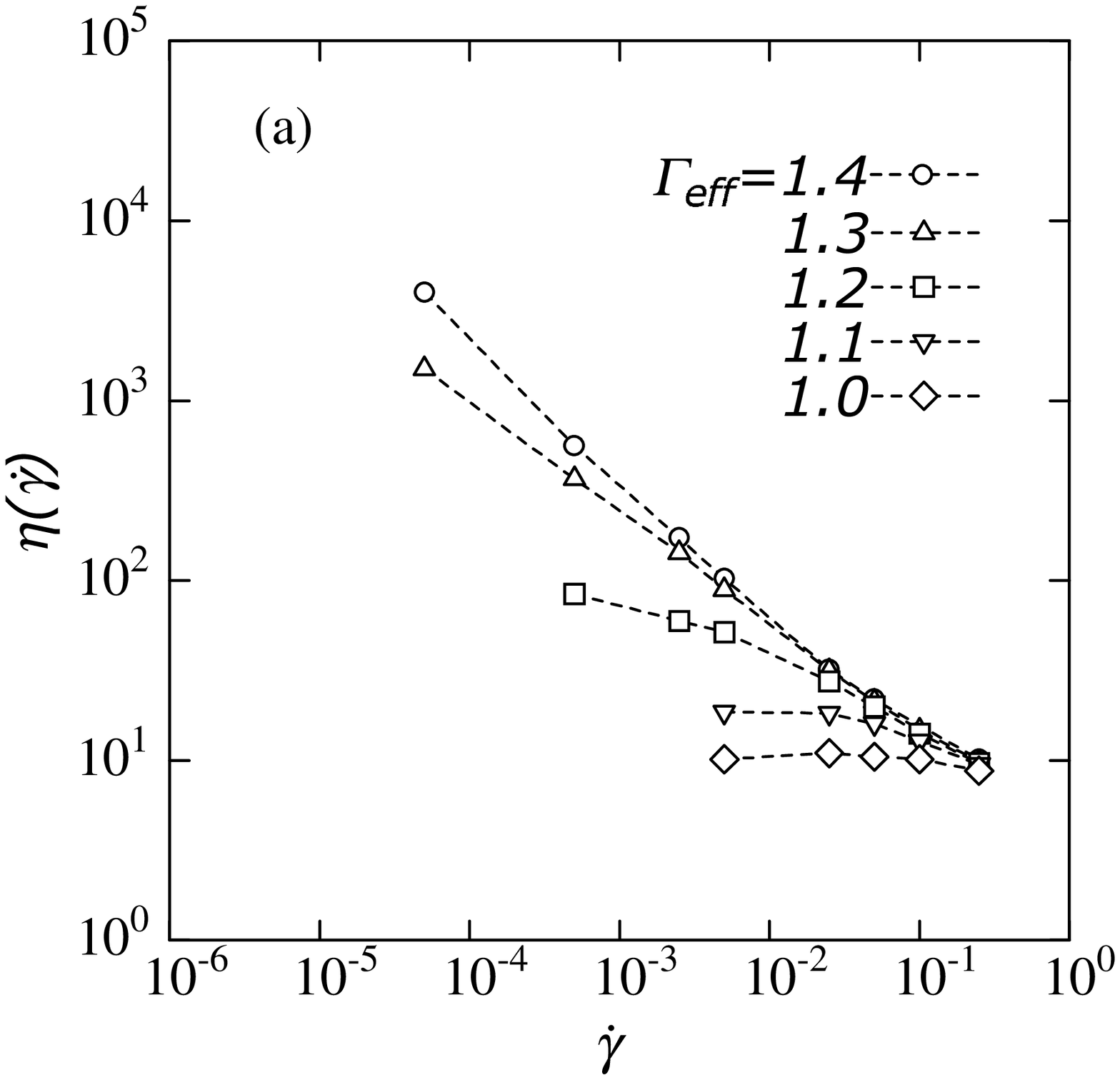}
\vspace{15mm}
\epsfxsize=3.0in
\epsfbox{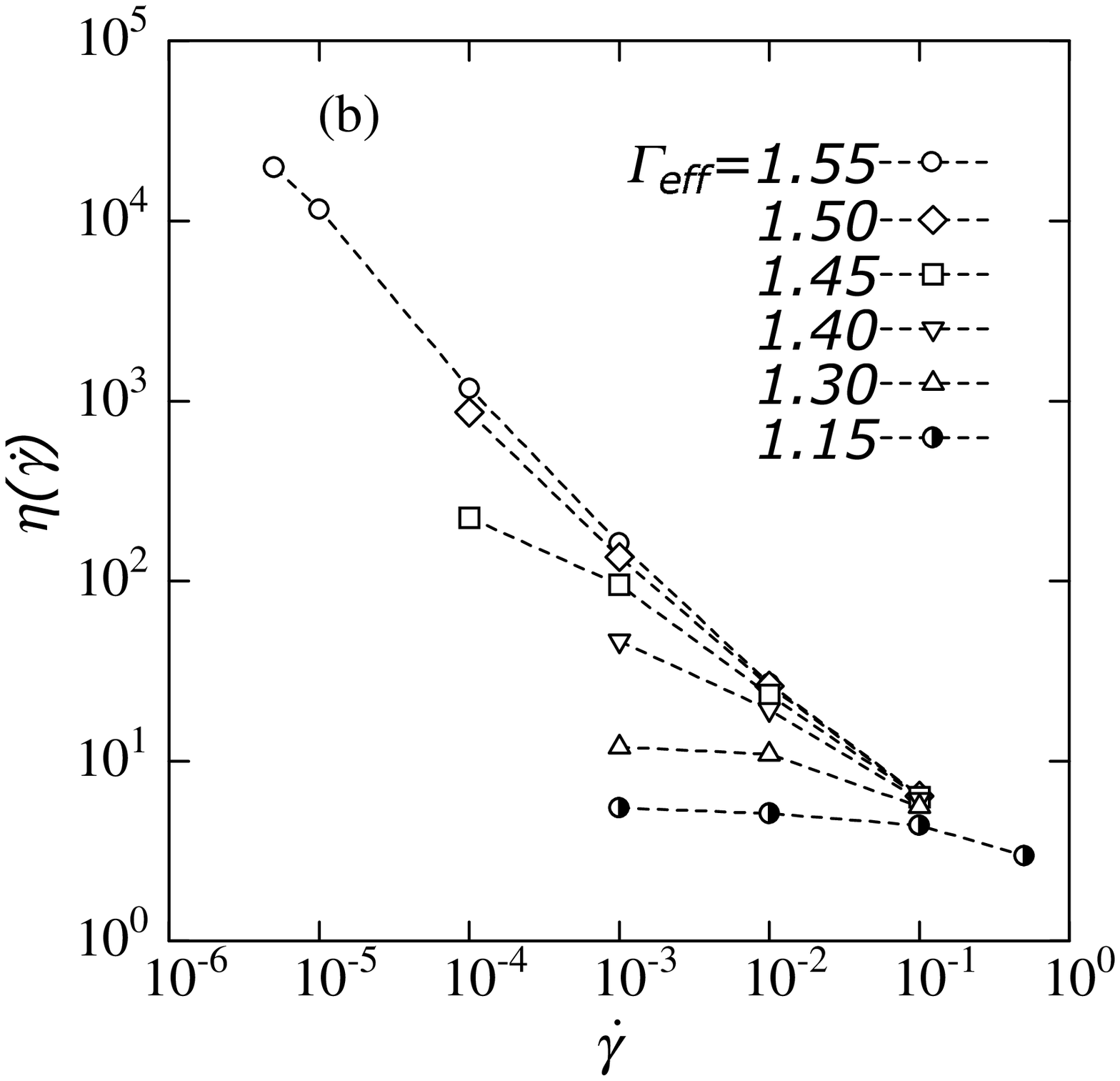}
\vspace{15mm}
\end{center}
\caption{\protect\narrowtext
The viscosity $\eta (\gdot)$
in units of  $\epsilon \tau_0/\sigma_1^d$
vesus the shear rate $\gdot$
in units of  $1/\tau_0$
  at various $\Gamma_{eff}$ in 2D (a) and 3D (b).
The data tend to become independent of
$\Ge$ at high shear.}
\vspace{10mm}
\end{figure}

%%%%%%%%%  Fig.12 %%%%%%%%%%%%
\begin{figure}[p]
\vspace*{10mm}
\begin{center}
\epsfxsize=3.0in
\epsfbox{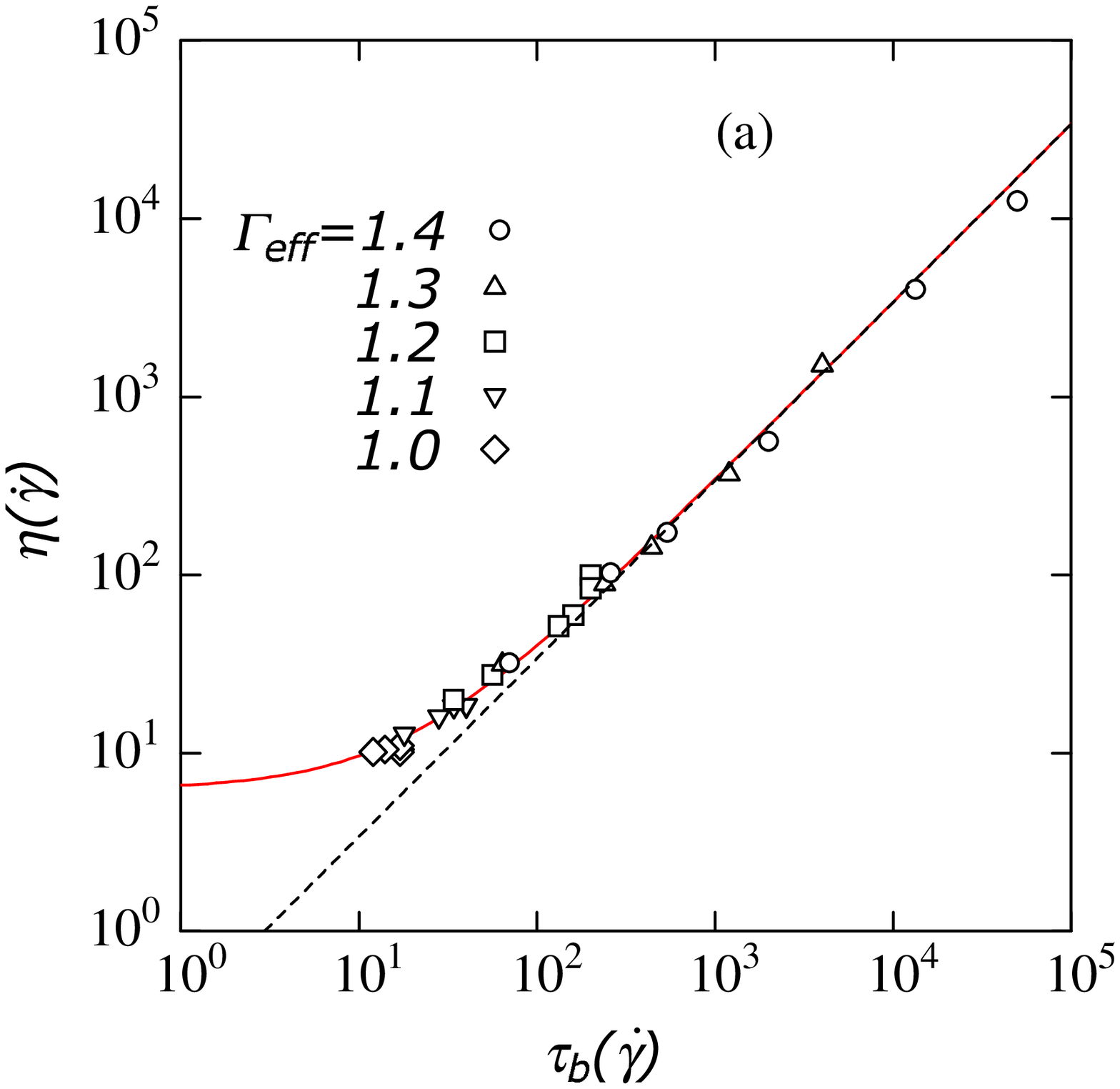}
\vspace{15mm}
\epsfxsize=3.0in
\epsfbox{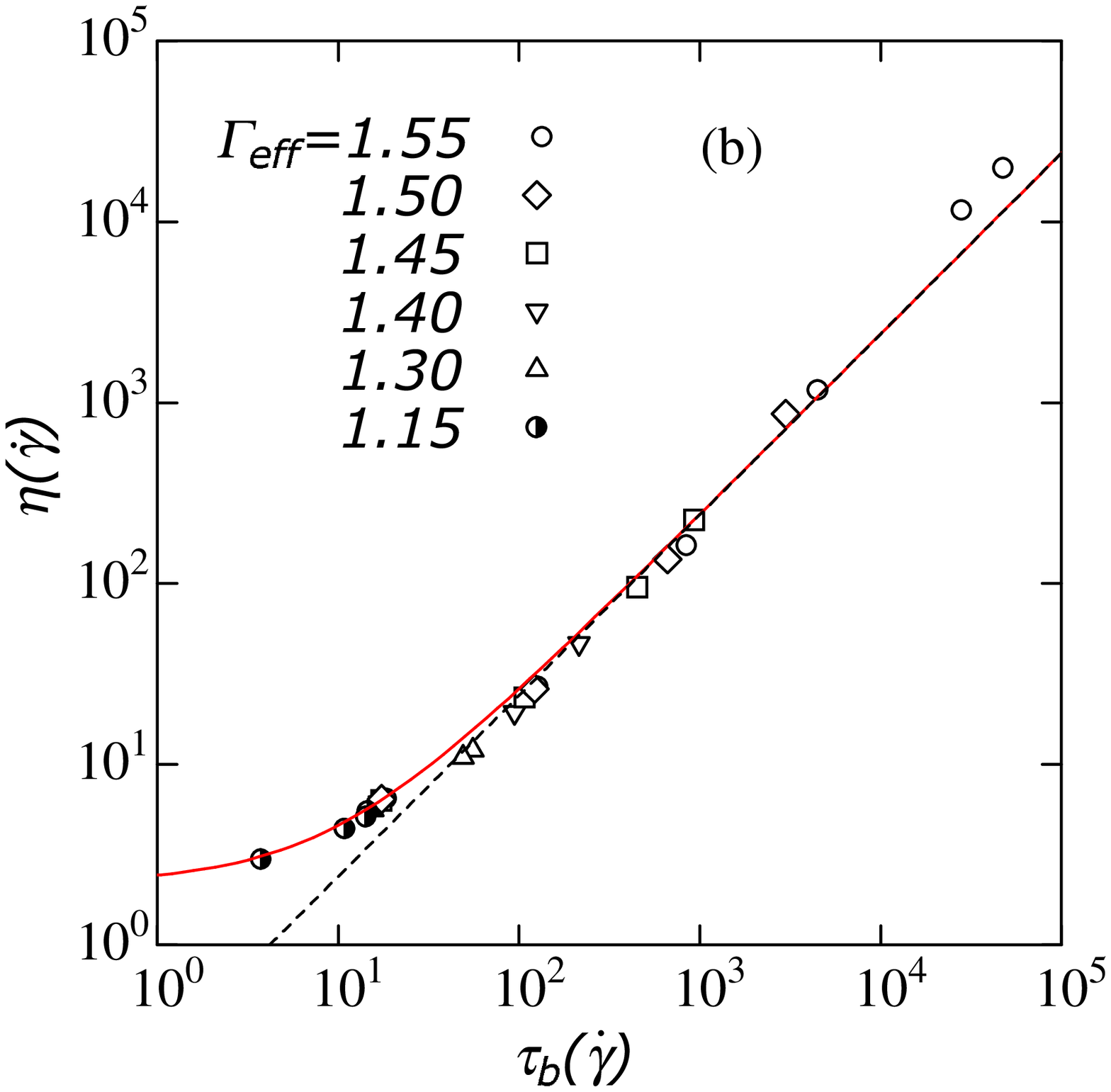}
\vspace{15mm}
\end{center}
\caption{\protect\narrowtext
$\eta (\gdot)$
vesus s $\tau_b(\gdot)$ for various $\Ge$
in 2D (a) and 3D (b). The
$\eta(\gdot)$ is determined by $\tau_b(\gdot)$ only
irrespectively of $\Ge$.
}
\vspace{10mm}
\end{figure}

%%%%%%%%%  Fig.13 %%%%%%%%%%%%
\begin{figure}[p]
\vspace*{10mm}
\begin{center}
\epsfxsize=3.0in
\epsfbox{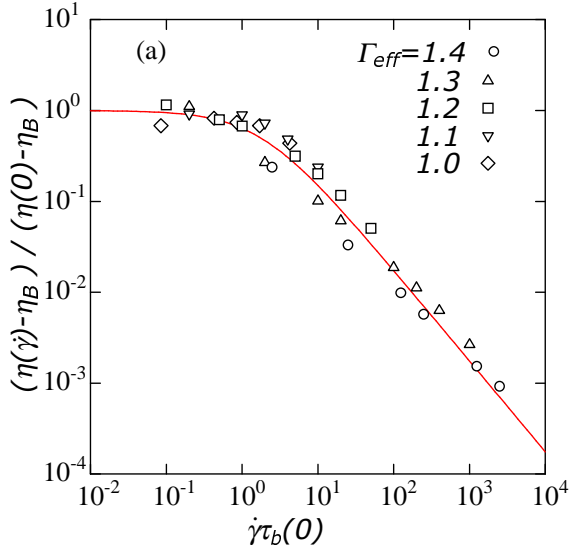}
\vspace{15mm}
\epsfxsize=3.0in
\epsfbox{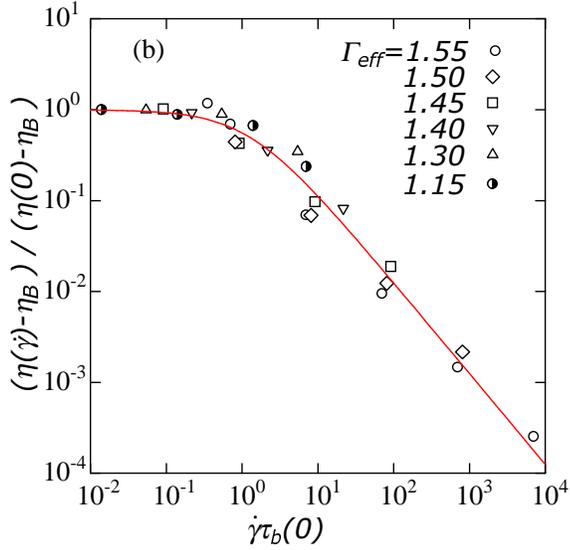}
\vspace{15mm}
\end{center}
\caption{\protect\narrowtext
$(\eta (\gdot) -\eta_B)/(\eta (0) -\eta_B)$
vs $\gdot \tau_b(0)$ in 2D (a) and 3D (b). The solid curve
is $1/(1+ A_b x)$ with $x=\gdot \tau_b(0)$. }
\vspace{10mm}
\end{figure}

%%%%%%%%%%%%% Fig.14 %%%%%%%%
\begin{figure}[p]
\begin{center}
\epsfxsize=3.0in
\epsfbox{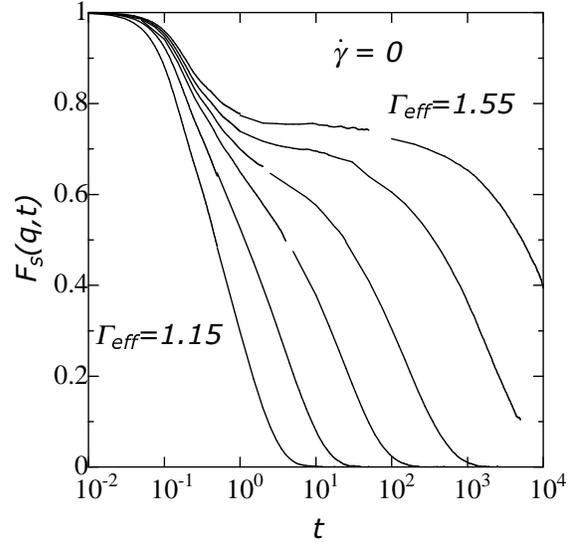}\vspace{-5mm}
\end{center}
\caption{\protect\narrowtext
The self part of the density-density
correlation function $F_s(q,t)$
at $q=2\pi$ and $\gdot=0$
 in 3D. $\Ge$ increases from left.
}
\vspace{10mm}
\end{figure}

%%%%%%%%%%%%% Fig.15  %%%%%%%%
\begin{figure}[p]
\begin{center}
\epsfxsize=3.0in
\epsfbox{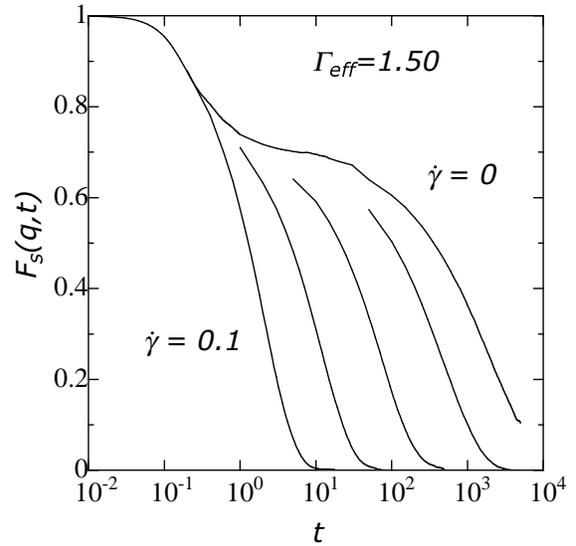}\vspace{-5mm}
\end{center}
\caption{\protect\narrowtext
The time correlation function $F_s(q,t)$ at
$q=2\pi$ defined by (6.1) and (6.2) in shear flow,
where $\gdot= 0$,
$10^{-4}$, $10^{-3}$, $10^{-2}$, $10^{-1}$  from
right.  The temperature  is fixed at $k_bT/\epsilon=0.267$
($\Gamma_{eff}=1.5$). Increasing $\gdot$ is equivalent
to raising $T$.
}
\vspace{10mm}
\end{figure}

%%%%%%%%%%%% Fig.16 %%%%%%%%
\begin{figure}[p]
\begin{center}
\epsfxsize=3.0in
\epsfbox{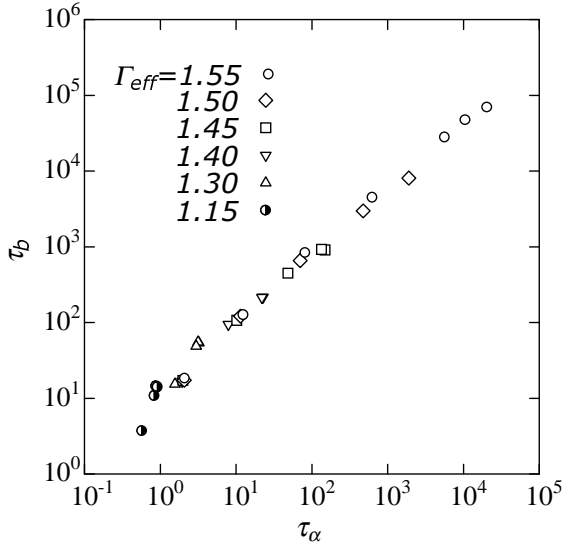}\vspace{-5mm}
\end{center}
\caption{\protect\narrowtext
 The linear relationship
between $\tau_{\alpha}$ and $\tau_b$
 for various
$\Gamma_{eff}$ and $\gdot$ in 3D.
The $\tau_b$ is determined from the bond breakage (3.14),
and  $\tau_{\alpha}$ from the decay of
the time correlation function (6.7).
}
\vspace{10mm}
\end{figure}

%%%%%%%%%%%% Fig.17 %%%%%%%%
\begin{figure}[p]
\begin{center}
\epsfxsize=3.0in
\epsfbox{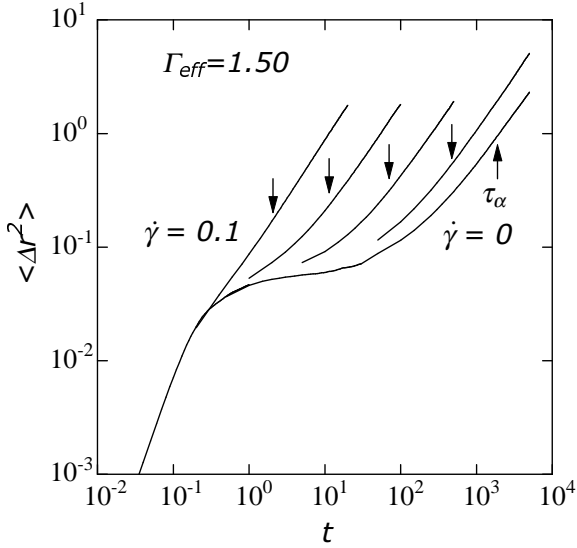}\vspace{-5mm}
\end{center}
\caption{\protect\narrowtext
The mean square displacement in sheared
states at $\Gamma_{eff}=1.5$.
The shear rate $\gdot$ is $0$, $10^{-4}$, $10^{-3}$,
$10^{-2}$, $10^{-1}$  from  right.  
Increasing $\gdot$ is equivalent to raising  $T$.
The arrows indicate $\tau_{\alpha}$ for each $\gdot$.
The diffusion (linear)
behavior is attained at $t \sim \tau_{\alpha}$.
}
\vspace{10mm}
\end{figure}

%%%%%%%%%%%% Fig.18 %%%%%%%%
\begin{figure}[p]
\begin{center}
\epsfxsize=3.0in
\epsfbox{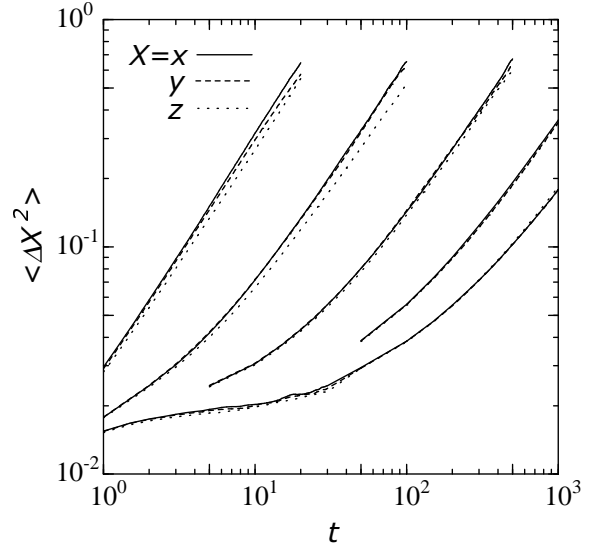}\vspace{-5mm}
\end{center}
\caption{\protect\narrowtext
The mean square displacements
of the $x$, $y$, and $z$ components.
They are very close to one another even
even in strong shear $\gdot \tau_b(0) \gg 1$.
This demonstrates surprising
isotropy of the distribution of the
displacement vector (6.2).
}
\vspace{10mm}
\end{figure}

%%%%%%%%%%%% Fig.19 %%%%%%%%
\begin{figure}
\begin{center}
\epsfxsize=3.0in
\epsfbox{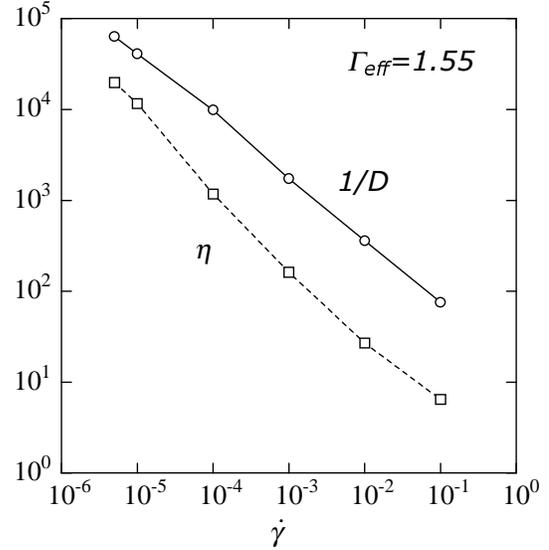}\vspace{-5mm}
\end{center}
\caption{\protect\narrowtext
Shear rate dependences of the
inverse diffusion constant $D$
and the viscosity $\eta$ at the lowest
temperature, $\Gamma_{eff}=1.55$.
The slope of $D^{-1}$ is noticeably smaller
than that of $\eta$. }
\vspace{10mm}
\end{figure}

%%%%%%%%%%%% Fig.20 %%%%%%%%
\begin{figure}
\begin{center}
\epsfxsize=3.0in
\epsfbox{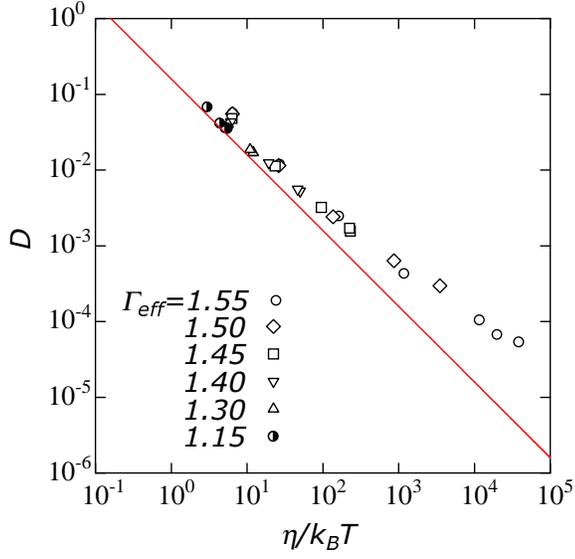}\vspace{-5mm}
\end{center}
\caption{\protect\narrowtext
The  diffusion
constant $D$ versus
the viscosity $\eta$ divided
by $k_BT$ for various $\Gamma_{eff}$
and $\dot{\gamma}$.
Here $D$ is measured in units of $\sigma_1^2\tau_0^{-1}$ and
$\eta/k_BT$
in units of  $\sigma_1^{-d} \tau_0$.
The solid line
represents  the Einstein-Stokes formula
$D =  k_BT/2\pi \eta \sigma_1$, which well
agrees with the numerical
data  for $\eta/k_BT \ls 10$.
}

\end{figure}

\clearpage

\end{multicols}

\end{document}